\documentclass[%
 reprint,
 amsmath,amssymb,
 aps,
]{revtex4-2}

\usepackage{graphicx}
\usepackage{dcolumn}
\usepackage{bm}
\usepackage{braket}
\usepackage{multirow}
\usepackage{hyperref}

\hypersetup{
    colorlinks=true,
    linkcolor=blue,
    filecolor=blue,      
    urlcolor=blue,
    citecolor=blue,
    pdfpagemode=FullScreen,
    }

\usepackage[english]{babel}
\begin{document}

\preprint{APS/123-QED}

\title{Continuous-variable state moments from randomized homodyne and heterodyne measurements}

\author{Michael Tsesmelis}
\affiliation{%
 Centre for Quantum Technologies, National University of Singapore, 117543 Singapore, Singapore
}%
\author{Moritz Straeter}%
\affiliation{%
 Centre for Quantum Technologies, National University of Singapore, 117543 Singapore, Singapore
}%
\author{Leong-Chuan Kwek}%
\affiliation{%
 Centre for Quantum Technologies, National University of Singapore, 117543 Singapore, Singapore
}%
\affiliation{%
MajuLab, CNRS-UNS-NUS-NTU International Joint Research Unit, UMI 3654, 117543 Singapore, Singapore
}%
\affiliation{%
 National Institute of Education, Nanyang Technological - University, 1 Nanyang Walk, 637616 Singapore, Singapore
}%

\date{\today}

\begin{abstract}
Continuous-variable (CV) quantum states are naturally characterized by their moments, defined as expectation values of products of single- or multimode ladder operators. Many CV Hamiltonians and quantum algorithms are formulated directly in terms of these moments, and therefore an efficient procedure to estimate moments with limited state measurements is necessary. In this paper, we present a protocol for shadow tomography of moment-generating functions (MGFs) of CV states based on randomized homodyne and heterodyne measurements. The resulting shadows enable an efficient and concurrent estimation of many multimode moments. Our complexity analysis shows that the number of measurements required to estimate a certain moment to a given precision grows exponentially in the order of the moments. Finally, we assess the precision of these moment estimators by applying them to two tasks: detecting entanglement in both Gaussian and non-Gaussian states via the Shchukin–Vogel protocol, and characterizing optical loss in a photonic chip. We demonstrate that both tasks can be accomplished with only a few thousand randomized measurements. The small number of required measurements, combined with efficient sample processing, makes our protocol applicable to a wide range of CV simulation tasks.
\end{abstract}

\maketitle
 

\section{\label{sec:level1}Introduction}
Characterizing highly entangled quantum states is a challenging task in quantum information science, as the number of parameters required to describe the state grows exponentially with system size.
This problem is even more severe for continuous-variable (CV) systems, whose probability distributions take on the form of continuous fields in phase space. Although tomography and representations of \textit{Gaussian states}---CV states described solely by first and second moments---are efficient, the number of state copies needed to reconstruct \textit{non-Gaussian states} scales exponentially with the number of modes \cite{mele_learning_2024}. Overcoming this scaling challenge is of utmost importance, since non-Gaussian states are essential for CV systems to achieve quantum advantage over classical information processing \cite{ra_non-gaussian_2020, adesso_continuous_2014, braunstein_quantum_2005, gomes_quantum_2009}. For example, non-Gaussian states are employed in quantum error-correction protocols and in schemes that encode qubits as superpositions of coherent states \cite{gottesman_encoding_2001, mari_positive_2012, ralph_quantum_2003}. 

Early proposed methods to reconstruct CV states rely on the inverse Radon transform, which reassembles the two-dimensional Wigner distribution in phase space from one dimensional marginal distributions sampled with homodyne measurements \cite{Leonhardt1995-lk}. Experimental reconstructions of the Wigner function is therefore only carried out for few modes \cite{barbieri_non-gaussianity_2010, molmer_non-gaussian_2006}, as reconstructing a many-mode entangled state becomes infeasible. Indeed, for $M$ modes and a phase space discretization into $N$ cells per dimension, the state description in phase space has size $N^{2M}$. \newline

There is a silver lining if we wish to extract only state observables rather than perform full state tomography. Recent studies on shadow tomography of qubit systems \cite{brydges_probing_2019,cieslinski_analysing_2024, elben_randomized_2022, elben_mixed-state_2020} demonstrate that expectation values can be obtained from many copies of a system without full state tomography, by averaging over randomized local measurement outcomes. These efforts on discrete systems have inspired similar initiatives for CV systems \cite{becker_classical_2024, gandhari_precision_2023}. These CV proposals focus on building unbiased estimators of the state $\rho$ expressed in the Fock state basis. However, since CV measurements such as homodyne and heterodyne detection sample the CV state in phase space, extensive postprocessing is required to recover the Fock-state representation. This postprocessing also struggles with high-Fock-number states. 

In this work, we therefore opt to construct phase space rather than Fock-space estimators of the state to avoid these issues. More specifically, we will focus on the construction of unbiased estimators of \textit{moment-generating functions} (MGF) of the state, which are more naturally derived from CV measurements. Complete knowledge of any MGF also gives us complete knowledge about the state $\rho$ \cite{Serafini2021-pd}. From the MGF we can in turn easily derive the moments of a quantum state, which characterize the distribution of the state in phase space. Since many Hamiltonians and quantum witnesses are written solely in terms of state moments, a reconstructed state MGF can be the most straightforward procedure to run quantum simulations or protocols on CV states, as it bypasses the reconstruction of the density matrix of $\rho$ entirely.\newline

As a first application of our shadow tomography protocol, we certify the entanglement of bipartite Gaussian and non-Gaussian CV states. Shchukin and Vogel (SV) \cite{shchukin_inseparability_2005} proposed an inseparability criterion for bipartite states which relies on an infinite number of state  moments. In the following work, we will test our MGF shadow tomography protocol by estimating the SV criterion and certifying the entanglement of Gaussian and non-Gaussian bipartite entangled states. In a second demonstration of our shadow tomography protocol, we describe a protocol to measure dissipation in a multimode Gaussian circuit. This protocol could be used for instance in experimental settings to characterize the photon loss on specific modes of a photonic chip. 

The paper will be organized as follows. In Section \ref{sec:theory}, we first expand on the theory behind CV systems and shadow tomography. Following that, in Section \ref{sec:methods} we describe how from randomized measurements in phase space we can efficiently construct MGF and moment estimators. We also give a sample complexity for these moment estimators. In Section \ref{sec:results}, we demonstrate the robustness of the shadow tomography for entanglement detection as well as optical dissipation characterization. Finally, we discuss the implications of our work and conclude in Section \ref{sec:conclusion}.

\section{Theory}
\label{sec:theory}
\subsection{Continuous-Variable systems}
\label{sec:cv}
Continuous-variable (CV) systems differ from regular qubit or qudit systems by their infinite-dimensional descriptions. Two different but equivalent representations exist for CV states. The density matrix of a CV system is commonly expressed in the Fock (energy eigenstate) basis ${\ket{n}}$, where $n$ denotes a Fock occupation number. In this formalism, any single-mode density operator is constructed as $\rho = \sum^\infty_{m, n=0} \rho_{m, n}\ket{m}\bra{n}$. The description is similar to qudit systems and only differs by the size of the state's Hilbert space, which is infinite-dimensional for a CV system. Secondly, the Fourier-Weyl relation \cite{Serafini2021-pd} provides an equivalent representation of the single-mode state in continuous phase space, where the state can be described, for example, over a complete basis of displacement operators $D_\alpha$. These displacement operators displace a single-mode vacuum state $\ket{\text{vac}}$ into a coherent state $\ket{\alpha}$ such that 

\begin{equation}
D_\alpha \ket{\text{vac}} = \ket{\alpha} \ .
\end{equation}

A coherent state $\ket{\alpha}$ is the eigenvector of an operator $ \hat{a}= (\hat{x} + i\hat{p})/\sqrt{2}$ and $ \alpha=(x + ip)/\sqrt{2}$ is the associated eigenvalue. The coherent state $\ket{\alpha}$ represents a Gaussian state centered around the point $(x, p)$ in phase space. Any operator on a single mode, including the density operator, can in turn be expressed in terms of this infinite basis set of displacement operators:

\begin{equation}
\hat{O} = \frac{1}{\pi}\int_{\mathbb{C}} \text{d}\alpha \, \text{Tr}\left[ \hat{D}_\alpha \hat{O} \right] \hat{D}_{-\alpha} \ .
\end{equation}

For a density operator $\hat{O}=\rho$, the characteristic function $\chi(\alpha) = \text{Tr}\left[ \hat{D}_\alpha \rho \right]$ is the weight of each displacement operator. We thus get

\begin{equation}
\label{eq:rho_chi_fourier}
    \rho = \frac{1}{\pi}\int_{\mathbb{C}} \text{d}\alpha \, \chi(\alpha) \hat{D}_{-\alpha} \ .
\end{equation}

Importantly, there exists a whole family of $s$-ordered characteristic functions such that $\chi_s(\alpha) = \text{Tr}[\hat{D}_\alpha \rho] e^{\frac{s}{2}|\alpha|^2}$. By taking the complex Fourier transform of a characteristic function, we get an $s$-ordered quasi-probability distribution of the state in phase space,

\begin{equation}
\label{eq:characteristic_function}
    W_s(\alpha) = \frac{1}{\pi^2} \int_{\mathbb{C}} d\beta \, e^{\alpha\beta^*-\alpha^* \beta} \chi_s(\beta) \, .
\end{equation}

These quasiprobability distributions, $W_s(\alpha)$, provide useful representations of quantum states because Gaussian measurements effectively sample from them. For $s=0$, we recover the Wigner distribution $W_0(\alpha)$ or $W_0(x, p)$ depending on the coordinate system used. Homodyne measurements \cite{leonhardt_realistic_1993} sample the rotated quadrature $\hat{x}_\theta = \cos\theta \, \hat{x} - \sin\theta \, \hat{p}$. The probability of sampling along this quadrature is simply the marginal probability
\begin{equation}
\bra{x_\theta}\rho\ket{x_\theta} =\frac{1}{2}\int^{+\infty}_{-\infty} \text{d}x_{\theta-\frac{\pi}{2}}W_0(x, p) \ ,
\end{equation}

\noindent or more simply put a Radon transform of the Wigner distribution, $P_\theta(x_\theta)$ \cite{wunsche_ordered_2000}. Alternatively, the Husimi-Q distribution $Q(\alpha) = \frac{1}{\pi} \bra{\alpha}\rho\ket{\alpha} \equiv W_{-1}(\alpha)$ describes the probability of finding the state $\rho$ in the coherent state $\ket{\alpha}$. Heterodyne measurements \cite{Serafini2021-pd} sample the state directly from $Q(\alpha)$. Finally, the Glauber-Sudarshan representation $P(\alpha) \equiv W_1(\alpha)$ helps us expand the quantum state $\rho$ in terms of an infinite basis of coherent states $\ket{\alpha}$

\begin{equation}
\rho = \int_\mathbb{C} \text{d}^2\alpha \, P(\alpha) \ket{\alpha}\bra{\alpha} \ .
\end{equation}

$P(\alpha)$ and $Q(\alpha)$ are not equivalent because coherent states are not orthogonal to each other, i.e. $\langle\alpha|\beta\rangle\neq0$. The Glauber-Sudarshan representation is difficult to sample from experimentally and therefore is not written as an expectation value the same way the Wigner or the Husimi-Q distribution are. The Glauber-Sudarshan representation is however necessary to find the final form of the normally-ordered moment estimators of our state. All three representations are parametrized by $M$ complex values for $M$ modes if the state describes a multimode system.

Different experimental realizations of CV quantum computing exist. Encoding a state $\rho$ into the Fock levels of a multimode optical system is a preferred method due to its relatively simple and scalable nature \cite{adesso_continuous_2014}. Other physical realizations of CV quantum computing include optomechanical  systems \cite{kanari-naish_two-mode_2022} and vibrational modes of trapped ions \cite{chen_quantum_2021}.
\subsection{Shadow Tomography}
\label{sec:tomography}

\begin{figure*}[htb]
\includegraphics[width=0.7\textwidth]{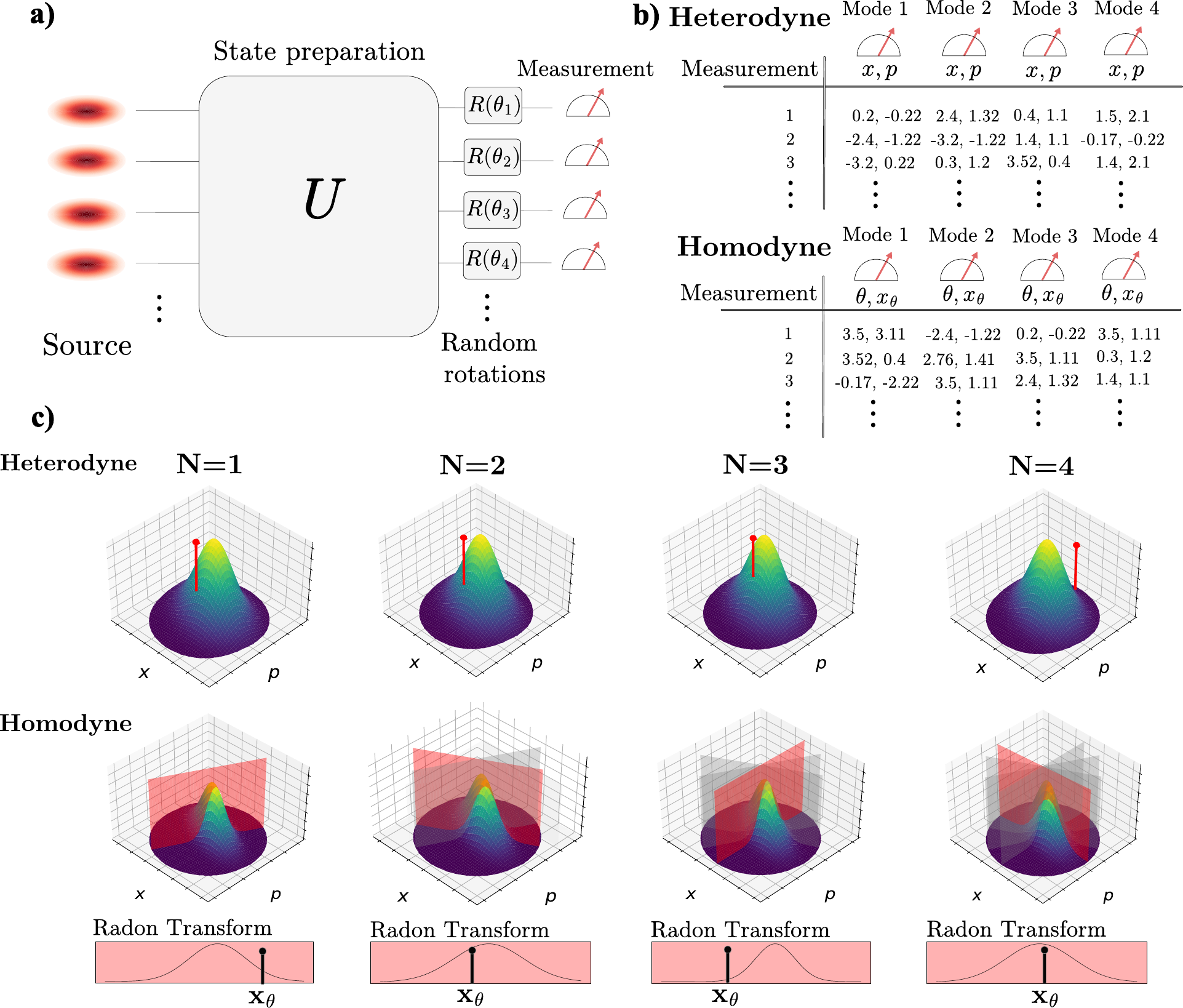}%
\caption{\label{fig:randomized_measurements}a) After preparing a state with some process $U$, random rotations are applied locally to each mode and the final state is measured. b-c) In the homodyne setting, the rotation $R(\theta)$ determines the measurement angle and the measurement device samples from the Radon-transformed Wigner function, obtaining $x_\theta$. On the other hand, heterodyne measurements directly sample points $(x, p)$ from the Husimi-Q quasiprobability distribution.}
\end{figure*}

Shadow tomography is a method used to reconstruct a quantum state or to approximate expectation values using a limited number of randomized measurements \cite{elben_randomized_2022, becker_classical_2024}. Randomized measurement protocols enable a \textit{measure first, analyze later} strategy; once all the measurements are collected, it is possible to derive from them many local observables up to a constant precision. In shadow tomography on qubits, it was also shown that the randomization process simplifies noise channels, making their effect predictable and easier to account for analytically in postprocessing \cite{koh_classical_2022, chen_robust_2021}. Traditionally, shadow tomography in CV systems has $N$ copies of a state and applies the following procedure: (i) apply local rotations parametrized by randomly generated rotation parameters $\bm{\mathsf{\theta}}$, (ii) measure the state, and finally (iii) undo the rotation and build a snapshot of the state, $\tilde{\rho}^{(i)}$ \cite{becker_classical_2024}. The measurement procedure for CV systems is described in Fig. \ref{fig:randomized_measurements}. Averaging over all $N$ snapshots yields an approximate reconstruction of the state, or shadow, $\tilde{\rho}$ \cite{becker_classical_2024}. 

In our paper, instead of reconstructing the state $\rho$, our intended goal is to construct an unbiased estimator of the characteristic function $\chi$. As we will show, it is possible to measure moments $\langle\prod_i^M \ \hat{a}_i^{\dagger k_i+l_i} \hat{a}_i^{k_i} \rangle$ directly from the characteristic functions, and therefore it is more straightforward to build shadows $\tilde{\chi}$ instead of $\tilde{\rho}$ if the observables of interest are moments only. Indeed, when proposing a protocol for the reconstruction $\tilde{\rho}$, Becker et al. \cite{becker_classical_2024} already proposed an estimator for the reconstructed characteristic function $\tilde{\chi}$ as an average over $N$ single-measurement heterodyne snapshots. After heterodyning an $M$-mode state, they obtain a set of measurements $\{\mathbf{x}^{(i)}\}^N_{i=1}$, where $\mathbf{x}^{(i)} = (x_0, p_0, \dots, x_M, p_M) \in \mathbb{R}^{2M}$ is the result of a single measurement run, and the reconstructed characteristic function is
\begin{equation}
\label{eq:shadow}
\tilde{\chi}(\mathbf{u}) = \frac{e^{\frac{\left\lVert\mathbf{u}\right\rVert^2}{4}}}{N} \sum^N_{i=1} e^{-i \mathbf{u}^T\Gamma \mathbf{x}^{(i)}} \ ,
\end{equation}

\noindent with $\mathbf{u} = (x'_0, p'_0, \dots, x'_M, p'_M) \in \mathbb{R}^{2M}$ the characteristic function coordinate and the symplectic matrix

\begin{equation}
\begin{aligned}
\Gamma &= \bigoplus^M_{j=1}\Gamma_1 \ , \\
\Gamma_1 &= \begin{pmatrix}0 & 1\\ -1 & 0\end{pmatrix} \ .
\end{aligned}
\end{equation}
The rotation parameters $\mathbf{\theta}$ do not appear in the reconstruction procedure for heterodyne measurements, as heterodyne measurements are rotationally invariant \cite{becker_classical_2024}. In the limit of infinite measurements $N \rightarrow \infty$, the normalized sum in Eq. \eqref{eq:shadow} becomes the expectation value

\begin{equation}
\label{eq:characteristic}
\begin{aligned}
e^{\frac{\left\lVert\mathbf{u}\right\rVert^2}{4}} \langle e^{-i \mathbf{u}^T\Gamma \mathbf{x}} \rangle_\mathbf{x} &= e^{\frac{\left\lVert\mathbf{u}\right\rVert^2}{4}} \int_{\mathbb{R}^{2M}} \text{d}\mathbf{x} \, Q(\mathbf{x}) \, e^{-i \mathbf{u}^T\Gamma \mathbf{x}} \\
&= e^{\frac{\left\lVert\mathbf{u}\right\rVert^2}{4}} \chi_{-1}(\mathbf{u}) = \chi_0(\mathbf{u})\ ,\\
\end{aligned}
\end{equation}

\noindent where in the first line we expand the expectation value over the distribution $Q(\mathbf{x})$ of heterodyne measurement samples. Becker et al.'s \cite{becker_classical_2024} procedure thus uses samples of the Husimi-Q function (and the associated $\chi_{-1}$ describing it) and from them reconstructs the characteristic function associated with the Wigner function, $\chi_0$. Becker et al. also recognize that the exponentially growing term in Eq. \eqref{eq:characteristic} leads to an integral that can fail to converge when plugged into Eq. \eqref{eq:rho_chi_fourier}. This makes it hard to reconstruct the density matrix $\rho$. Their solution to this is to limit the use of their shadows to the calculation of expectations values of operators $\hat{O}$ which suppress the exponential growth of the characteristic function at large $|\mathbf{u}|$. This limitation partly motivates the present work. Instead of reconstructing an approximation to the density matrix according to Eq. \eqref{eq:rho_chi_fourier}, our work instead stops at the reconstructed characteristic function in Eq. \eqref{eq:shadow}, and shows how this object is sufficient to derive many important properties of the state.

In the homodyne setting, this work is motivated by a different kind of issue; the classical shadow of the characteristic function as derived in \cite{becker_classical_2024} is defined only in the distributional limit. For a single mode shadow built from $N$ measurements, we have

\begin{equation}
\label{eq:becker_homodyne}
    \tilde{\chi}(u) = \frac{2\pi}{N} \sum^N_{i=1} \left\lVert u \right\rVert \delta((S(\theta^{(i)})u)_2) e^{-iu^T\Gamma S^{-1}(\theta^{(i)})x^{(i)}} \ ,
\end{equation}

\noindent where the matrix $S(\theta)$ rotates the quadrature of the homodyne measurement. Operating directly on the characteristic function is difficult due to the delta function. Our work solves this problem by defining a new MGF which avoids the distributional singularities present in the characteristic function, and which can thus be operated on more easily to derive the state moments.

\section{Characteristic Function Shadows}
\label{sec:methods}
Our intended goal is to estimate the expectation value of normally-ordered operators $\langle:\prod_i \ \hat{a}_i^{\dagger k_i+l_i} \hat{a}_i^{k_i}: \rangle$ for each mode $i$ of a CV multi-mode system. Any moment can then be obtained by reordering using the canonical commutation relation for bosonic ladder operators. As we will see below, for heterodyne measurements it is possible to calculate the expectation value of the normal moment operators directly from derivatives of the characteristic function. For homodyne measurements, \cite{becker_classical_2024} defines the  characteristic function in the distributional limit only as shown in Eq. \eqref{eq:becker_homodyne}, and thus we must first define a new MGF of the state. The starting point of both methods are the seminal works of Cahill and Glauber on boson amplitude operators \cite{cahill_density_1969, cahill_ordered_1969}, which relate moments to characteristic functions.

\subsection{Heterodyne Detection}

The normally-ordered moment operators can be obtained from the derivatives of $\chi_1$ \cite{Serafini2021-pd}, the characteristic function associated to the Glauber-Sudarshan distribution, as

\begin{equation}
\langle \hat{a}^{\dagger m}\hat{a}^n \rangle_1 = \left.\left( \frac{\partial}{\partial \alpha}\right)^m \left( -\frac{\partial}{\partial \alpha^*}\right)^n\chi_{1}(\alpha) \right|_{\alpha=0} \, .
\end{equation}

As explained in \ref{sec:cv}, the issue is that the Glauber-Sudarshan distribution $W_1$ cannot be sampled from directly, which means it is not possible to obtain $\chi_1$ by complex Fourier Transform of a CV measurement only. Instead, we sample first from $W_{-1}$ using heterodyne measurements, and after a complex Fourier transform, we obtain $\chi_{-1}$. We can then use the relationship between different $s$-ordered characteristic functions in Sec. \ref{sec:cv} to derive the expected moments of normally-ordered operators for $M$-mode systems  \cite{Serafini2021-pd}:

\begin{widetext}
\begin{align}
\label{eq:operators_from_derivative}
\left\langle \prod_{j=1}^M \hat{a}_j^{\dagger m_j} \hat{a}_j^{n_j} \right\rangle_1
&=  \prod_{j=1}^M \left( \frac{\partial}{\partial \alpha_j} \right)^{m_j}
\left( -\frac{\partial}{\partial \alpha_j^*} \right)^{n_j}
\left.\left( \chi_{-1}(\boldsymbol{\alpha}) \, e^{ \sum_{k=1}^M |\alpha_k|^2} \right) \right| _{\boldsymbol{\alpha} = 0} \ .
\end{align}  
\end{widetext}

In our entanglement detection scheme, we will look specifically at the two-mode system where $M=2$. The derivatives with regards to a complex variable $\alpha = x+iy$ require Wirtinger calculus. For more details regarding the derivatives with respect to complex variables, please refer to the Supplementary Information. Complex derivatives rely on derivatives along different axes of the many-body phase space. If we store the values of the characteristic function around the phase space point $\mathbf{\alpha} = \mathbf{0}$, we can use numerical differentiation methods to calculate any gradient in (x-p) phase space and in turn any gradient of $\chi$ with regards to $\alpha$ or $\alpha^*$. For estimates of the precision of finite-difference approximations to complex derivatives, see Ref.~\cite{fornberg_finite_2022}. 

\subsection{Homodyne Detection}

Estimating moments with derivatives of the characteristic function faces a severe challenge with homodyne measurements. In Eq. \eqref{eq:becker_homodyne}, we showed how samples of the marginal Wigner distribution cannot be used to reconstruct a two-dimensional characteristic function. Indeed, the rotation of the homodyne quadrature is controlled by the matrix $S$ and the characteristic function is non-zero only along that same quadrature angle. More succinctly, the relation between the homodyne marginals of the Wigner distribution (also known as a Radon-transformed Wigner) $P_\theta( x_\theta)$ and its characteristic function at angle $\theta$ and radius $u_\theta$ is simply \cite{Leonhardt1995-lk}

\begin{equation}
\chi_0(\theta, u_\theta) = \int^\infty_{-\infty} \text{d}x_\theta \, P_\theta(x_\theta) \, e^{-iu_\theta x_\theta} \ .
\end{equation}

\begin{figure}[htb]
\includegraphics[width=1.0\columnwidth]{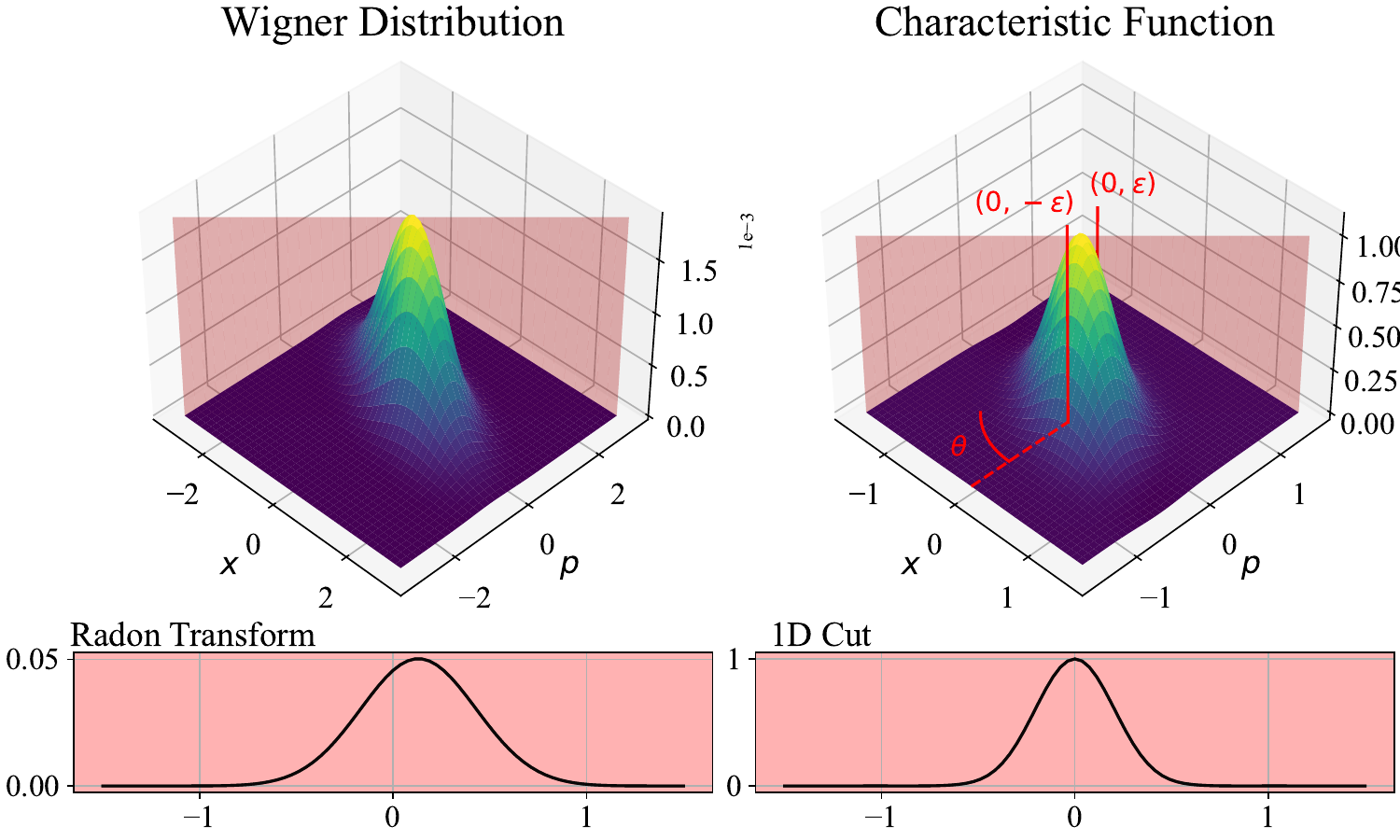}
\caption{\label{fig:fourier} The characteristic function $\chi(\theta, u_\theta)$ of a state along some angle $\theta$ is the complex Fourier Transform of a marginal probability distribution of the Radon-transformed Wigner function $P_\theta( x_\theta)$ along the same quadrature angle $\theta$ of the phase space. Homodyne measurements sample directly from this marginal distribution function. For a given angle, we can therefore only rebuild the characteristic function from Wigner samples along a specific angle $\theta$. However, determining the expectation value of normally-ordered operators requires knowledge about the characteristic function at well-defined points in phase space $(x, p)$, which usually do not lie on the same angle $\theta$. For instance, when evaluating $\langle\hat{a}^\dagger\rangle$ and the derivative $\frac{\partial \chi_1}{\partial \alpha} = \frac{1}{2}\left(\frac{\partial \chi_1}{\partial x} - i\frac{\partial \chi_1}{\partial p} \right)$, the second term on the right-hand side requires the value of $\chi$ at phase-space points $(0, \epsilon)$ and $(0, -\epsilon)$ with $\epsilon \ll 1$ for numerical differentiation. These two points are depicted as two dark-red projection lines in the figure above. These points are not captured by the characteristic function at slice $\theta$, and thus any measurement made along that quadrature angle does not contribute to estimating the value of the characteristic function at the required points. Most randomized measurements in a homodyne setting will therefore not contribute to describing the characteristic function at the required points $(\theta, u_\theta)$. The expectation values are taken as an average over less points, thus making the expectation values less precise.}
\end{figure}

Thus, the marginal Wigner distribution along a certain homodyne quadrature characterized by angle $\theta$ only provides information about the characteristic function along that same angle, meaning for each randomized measurement the snapshot $\tilde{\chi}^{(i)}_0$ has support on a one-dimensional slice rather than the full two-dimensional phase space \cite{Leonhardt1995-lk, becker_classical_2024}. Fig. \ref{fig:fourier} is a graphical description of this relationship. For a set of randomized angles $\{\theta_i\}$, the final estimator $\tilde{\chi}_0(\mathbf{u})$ is thus sparsely defined only along angles in $\{\theta_i\}$. The characteristic function at each angle $\theta_i$ is also taken as an average over less measurements, since only homodyne measurements at the same angle $\theta_i$ contribute. Historically, one approach to circumvent this estimator-construction problem was to sample the Wigner function at equidistant measurement angles in order to determine the expectation value of a given function of the ladder operators $\hat{a}, \hat{a}^\dagger$ \cite{wunsche_ordered_2000}. This solution however bypasses the advantage of randomized measurements, which as their name suggests should occur at arbitrarily chosen angles. Ultimately, we need to construct a representation of a (possibly entangled, multimode) quantum state exhibiting certain key features --- the representation (i) is suited for fast moment derivations, (ii) has limited memory requirements and finally (iii) is suited for randomized sampling. Since the characteristic function $\chi$ is not suited for randomized homodyne samples due to the issues mentioned above, we instead devise a new MGF to replace $\chi$ which will be used in the homodyne setting.

The starting point of this search for a new MGF comes from Richter \cite{richter_realistic_2000}, who introduced a different single-mode generalized MGF,

\begin{equation}
    \begin{aligned}
    M_l(\mu) \equiv \text{Tr}\left[\rho \hat{a}^{\dagger l} (1-\mu)^{\hat{a}^\dagger \hat{a}}\right] \ ,
    \end{aligned}
\end{equation}

\noindent where the factor $(1-\mu)^{\hat{a}^\dagger \hat{a}}$ acts as a photon-number–dependent weight, allowing access to higher-order normally-ordered moments.

When $l=0$, we get back an MGF widely used in describing photon statistics \cite{glauber1965, barnett_statistics_2018}. Its derivatives $ \left.(\text{d}/\text{d}\mu)^k M_l(\mu) \right|_{\mu=1}$ are related to the density matrix elements $\rho_{k, k+l}$. Finally and more importantly for us, using this MGF, we can derive a set of moments

\begin{equation}
\label{eq:moment_from_M}
    \begin{aligned}
    \langle \hat{a}^{\dagger k+l} \hat{a}^{k} \rangle = (-1)^k \frac{d^k}{d\mu^k}M_l(\mu)|_{\mu=0} \ .
    \end{aligned}
\end{equation}

Richter also provided a method to derive $M_l$ from the full sinogram $P_\theta(x_\theta)$ \footnote{A sinogram is a description of the state as Radon transformations of the state over a continuous angle $\theta \in [0, \pi)$.} of the Wigner probability distribution,

\begin{equation}
\label{eq:M_generating_function}
    \begin{aligned}
    M_l(\mu) = \int^\infty_{-\infty} dx_\theta  \int^{2\pi}_0d\theta \,P_\theta( x_\theta) \, S_M(x_\theta, \mu, l) e^{-il\theta} \ ,
    \end{aligned}
\end{equation}

\noindent where $S_M(x_\theta, \mu, l)$ is a sampling function defined as  

\begin{equation}
    \label{eq:SM}
    \begin{aligned}
    S_M(x_\theta, \mu, l) = \frac{\sqrt{l!}}{2\pi(2-\mu)}\left[\frac{1}{(2-\mu)\mu}\right]^{l/2} f_{0,l}\left(\sqrt{\frac{\mu}{2-\mu}} x_\theta\right) \ .
    \end{aligned}
\end{equation}

\noindent with pattern functions $f_{0, l}$ (for more information about pattern functions, please refer to the Supplementary Information). Importantly, each randomized measurement sample $(\theta, x_\theta)$ contribues to the full range of $l$ and $\mu$ parametrizing $M_{l}(\mu)$, see Eq. \eqref{eq:M_generating_function}. This means that all homodyne measurements contribute to the estimation of $M_{l}(\mu)$ at any $l, \mu$.

There are issues with the current form of $M_l(\mu)$ however. First, getting derivatives at $\mu=0$ according to Eq. \eqref{eq:moment_from_M} is  complicated by the singularity at the origin. Also, pattern functions $f_{0,l}$ become increasingly difficult to compute with larger $l$. Instead, we derive a new expression for $S_M(x_\theta, \mu, l)$ thanks to the expression of a single-mode moment $\langle \hat{a}^{\dagger k+l} \hat{a}^k\rangle$ introduced in \cite{wunsche_ordered_2000},

\begin{equation}
    \label{eq:moment_from_hermite}
    \begin{aligned}
    \langle \hat{a}^{\dagger k+l} \hat{a}^{k} \rangle = \left[ \pi \sqrt{2^{2k+l}} \binom{2k+l}{k+l}\right]^{-1} \int^{\pi}_0 d\theta \\ \times \int^{+\infty}_{-\infty} dx_\theta \, P_\theta(x_\theta) \,H_{2k+l}(x_\theta)e^{-il\theta} \ .
    \end{aligned}
\end{equation}

By equating the right-hand sides of Eqs. \eqref{eq:moment_from_M} and \eqref{eq:moment_from_hermite}, we can derive an easy representation of the sampling function,

\begin{equation}
    \begin{aligned}
    \label{eq:moment_equating}
    S_M(x_\theta, \mu, l) = &\sum^\infty_{k=0}\frac{\mu^k}{k!}\left[ (-1)^k  \pi\sqrt{2^{2k+l}} \binom{2k+l}{k+l}\right]^{-1} \\&\times H_{2k+l}(x_\theta) \ .
    \end{aligned}
\end{equation}

The sum in $S_M$ can be cut off at $k=L$, where $L$ is the highest moment that we want to evaluate using the shadow tomography protocol. We now have a new expression for the MGF, which we will denote by $\mathcal{M}_l(\mu)$. The single-shot estimator of this new MGF is therefore

\begin{equation}
\label{eq:single_shot_estimator}
\begin{aligned}
\mathcal{M}_l^{(i)}(\mu) =& \sum^L_{k=0}\frac{\mu^k}{k!}\left[ (-1)^k  \pi\sqrt{2^{2k+l}} \binom{2k+l}{k+l}\right]^{-1} \\&\times H_{2k+l}(x_{\theta^{(i)}})e^{-i l \theta^{(i)}} \, .
\end{aligned}
\end{equation}

We see that this single-shot estimator does not rely on the pattern functions and is extremely fast at processing the measurements, since we must only multiply the homodyne samples by some simple factors such as Hermite polynomials. Overall, the postprocessing for the MGF $\mathcal{M}_l$ is thus basically free. 

In a multimode setting, local measurements on all modes yield a product state after projection, i.e. $\rho^{(i)} = \otimes^M_{m=1} \rho^{(i)}_m$. In the bipartite scenario, the generating function of such a product state, $\mathcal{M}_{l_1, l_2}(\mu_1, \mu_2)$, is 

\begin{equation}
    \label{eq:proof_separable_state}
    \begin{aligned}
        \mathcal{M}_{l_1, l_2}(\mu_1, \mu_2) &= \text{Tr}(\rho (\hat{a}^{\dagger l_1}(1-\mu_1)^{\hat{a}^\dagger\hat{a}} \otimes \hat{b}^{\dagger l_2}(1-\mu_2)^{\hat{b}^\dagger\hat{b}})) \\
        &= \text{Tr}(\rho_1 \hat{a}^{\dagger l_1}(1-\mu_1)^{\hat{a}^\dagger\hat{a}})\text{Tr}(\rho_2 \hat{b}^{\dagger l_2}(1-\mu_2)^{\hat{b}^\dagger\hat{b}})\\
        &= \mathcal{M}_{l_1}(\mu_1) \times \mathcal{M}_{l_2}(\mu_2) \ .
    \end{aligned}
\end{equation}

\noindent Therefore, for a two-mode product state, the MGF is simply the product of the generating functions of each subsystem. This relationship also generalizes to a product state of any number of modes. Next, we obtain the two-mode MGF shadow  $\widetilde{\mathcal{M}}_{l_1, l_2}(\mu_1, \mu_2)$ from an average over $N$ measurements, each yielding a product of single-mode MGF snapshots, 

\begin{equation}
\label{eq:two_mode_shadow}
\begin{aligned}
    \widetilde{\mathcal{M}}_{l_1, l_2}(\mu_1, \mu_2) = \frac{1}{N}\sum_{i=1}^N \mathcal{M}^{(i)}_{l_1}(\mu_1)\,\mathcal{M}^{(i)}_{l_2}(\mu_2) \ .
\end{aligned}
\end{equation}

In our randomized measurement protocol, instead of taking derivatives on the multimode MGF shadow  $\widetilde{\mathcal{M}}_{l_1, l_2}$, we can also derive the moments as partial derivatives of the product $\mathcal{M}^{(i)}_{l_1}(\mu_1)\mathcal{M}^{(i)}_{l_2}(\mu_2)$:

\begin{equation}
\begin{aligned}
    \label{eq:two_mode_derivatives}
    \langle \hat{a}^{\dagger k_1+l_1} &\hat{a}^{k_1} \hat{b}^{\dagger k_2+l_2} \hat{b}^{k_2} \rangle \\&= \frac{1}{N}\sum_{i=1}^N \left(\frac{d^{k_1}}{d\mu_1^{k_1}} \mathcal{\mathcal{M}}^{(i)}_{l_1}(\mu_1)\right)\left(\frac{d^{k_2}}{d\mu_2^{k_2}}\mathcal{M}^{(i)}_{l_2}(\mu_2)\right) \ .
\end{aligned}
\end{equation}

\noindent One advantage of classical shadows also becomes clear in this relationship. The shadow $\widetilde{\mathcal{M}}_{l_1, l_2}(\mu_1, \mu_2)$ on the left-hand side of Eq. \eqref{eq:two_mode_shadow} requires a memory footprint that scales exponentially in $M$. Instead, in postprocessing, it is more interesting to operate on the factorized representation $\prod^M_{m=1} \mathcal{M}^{(i)}_{l_m}(\mu_m)$, whose memory footprint scales linearly in $M$. 

In summary, we have introduced a new MGF $\mathcal{M}_{l}(\mu)$, which solves an important issue in the construction of characteristic function shadow $\tilde{\chi}$ for homodyne measurements and which does not require pattern functions $f_{0,l}$. In the next section, we will analyze the scaling complexity of using $\chi$ and $\mathcal{M}_l$ to derive normally-ordered moments of a state.


\subsection{Sampling efficiency}

When estimating a single-mode moment $\langle \hat{a}^{\dagger k+l} \hat{a}^k \rangle$ in the homodyne setting (as an example), we construct a random variable $Z_i$ from each measurement outcome as

\begin{equation}
Z_i = \frac{d^{k}}{d\mu^{k}} \mathcal{M}^{(i)}_{l}(\mu)  \, .
\end{equation}
The estimator is given by $\hat{Z}=\frac{1}{N}\sum^N Z_i$ with $N$ samples. Bernstein's inequality \cite{rebeschini_bernsteins_nodate} states that the sampling efficiency of this estimator depends on the variance $\operatorname{Var}[Z]$ of the samples $Z_i$, as well as the bound $R$ such that $|Z_i-\mathbb{E}[Z]|\leq R$. We would like to determine how many measurements are needed to attain maximal distance of $\epsilon$ between the estimator value $\hat{Z}$ and the exact expected value $\mathbb{E}[Z]$, with some probability $\mathbb{P}(X) \geq 1-\delta$:  

\begin{equation}
\mathbb{P}\Bigg( \left|\frac{1}{N} \sum_{i=1}^N \hat{Z}_i - \mathbb{E}[Z] \right| \leq \epsilon \Bigg) \geq 1 - \delta \, .
\end{equation}

Bernstein's inequality gives us this probability as

\begin{equation}
\mathbb{P}\Bigg( \left|\frac{1}{N} \sum_{i=1}^N \hat{Z}_i - \mathbb{E}[Z] \right| \leq \epsilon \Bigg) \leq 2 \, \mathrm{exp}\left(-\frac{N \epsilon^2}{2(\operatorname{Var}[Z]+ R\epsilon/3)} \right) \, .
\end{equation}

Together, these two inequalities tell us that the required number of samples is
\begin{equation}
\label{eq:measurement_number}
N \geq \ln\left(\frac{2}{\delta}\right) \frac{2\operatorname{Var}[Z]}{\epsilon^2} + \frac{2}{3} \frac{R}{\epsilon}.
\end{equation}

This procedure applies to the heterodyne setting as well. Next, we determine analytically both the variance of the estimator $\operatorname{Var}[Z]$ and the maximal distance from the expected value $R= \underset{\hat{Z}}{\operatorname{max}}\{|\hat{Z} - \mathbb{E}[Z]|\}$ for a given moment using our randomized measurement protocol for either heterodyne or homodyne measurements. We consider a CV state $\rho = \sum^D_{m, n=0} \rho_{m, n}\ket{m}\bra{n}$, with $D$ a Fock truncation.

\subsubsection{Heterodyne measurements}
In the case of heterodyne measurements, using Eqs. \eqref{eq:characteristic_function} and \eqref{eq:operators_from_derivative}, we can describe the single-mode characteristic function $\chi_1$ in terms of the Husimi-Q function as

\begin{equation}
\chi_1(\beta) = \int_\mathbb{C} d^2\alpha \, e^{(\alpha\beta^*-\alpha^*\beta)}\, Q(\alpha) \, e^{|\beta|^2} \, .
\end{equation}

The normally-ordered operators for a single mode are then obtained from the derivatives of $\chi_1(\beta)$

\begin{equation}
\begin{aligned}
\label{eq:kernel}
\langle \hat{a}^{\dagger k+l}\hat{a}^k\rangle &= \left.\left(\frac{\partial}{\partial\beta}\right)^{k+l} \left(-\frac{\partial}{\partial\beta^*}\right)^{k}\chi_1(\beta)\right|_{\beta=0}\\
&= \int_\mathbb{C}d^2 \alpha \, K_{k, l}(\alpha) \,Q(\alpha)
\end{aligned}
\end{equation}

\noindent where the kernel function is defined as

\begin{equation}
K_{k, l}(\alpha) = \left.\left(\frac{\partial}{\partial\beta}\right)^{k+l} \left(-\frac{\partial}{\partial\beta^*}\right)^{k} e^{\alpha \beta^*-\alpha^*\beta+\beta\beta^*}\right|_{\beta=0} .  
\end{equation}

Since $K_{k, l}(\alpha)$ is a polynomial in $\alpha$, $\alpha^*$ and $Q(\alpha)$ is a distribution, the variance of the heterodyne estimator can be upper bounded as 

\begin{equation}
\text{Var}^{(\mathrm{het})}[Z] \leq \int_\mathbb{C}d^2 \alpha \, |K_{k, l}(\alpha)|^2 \,Q(\alpha) = \langle Z^2\rangle \, .
\end{equation}

Expanding the squared kernel as

\begin{equation}
|K_{k, l}(\alpha)|^2=\sum_{q, r}\gamma_{q,r}\alpha^q(\alpha^*)^r ,
\end{equation}

\noindent each monomial contributes

\begin{equation}
\gamma_{q,r}\sum^D_{n=1} \rho_{n-r+q, n}\frac{(n+q)!}{\sqrt{(n-r+q)!\,n!}} 
\end{equation}

\noindent to the variance. Summing over all monomials gives us

\begin{equation}
\begin{aligned}
\operatorname{Var}^{(\mathrm{het})}[Z] \leq \sum_{q, r}\gamma_{q,r}\sum^D_{n=1} \rho_{n-r+q, n}\frac{(n+q)!}{\sqrt{(n-r+q)!\,n!}}  \,  .
\end{aligned}
\end{equation}

The factorial term satisfies
\begin{equation}
\frac{(n+q)!}{\sqrt{(n-r+q)!\,n!}} = n^{\frac{q+r}{2}}\left(1 + \mathcal{O}(n^{-1})\right) , \hspace{0.5cm} n\rightarrow\infty\, .
\end{equation}

The asymptotic behavior of the upper bound of $\operatorname{Var}^{(\mathrm{het})}[Z]$ is determined by the terms where $r=q=2k+l$, for which $\gamma_{q, r}$ is always positive because it is the coefficient of the highest-order monomial in $|K_{k,l}(\alpha)|^2$. Finally, our analysis shows that the second moment $\langle |Z|^2\rangle$ admits the asymptotic bound

\begin{equation}
\operatorname{Var}^{(\mathrm{het})}[Z] = \mathcal{O}\left(D^{2k+l}\right) \, .
\end{equation}

 This is an important discovery. Had the variance been exponential in the Fock level $n$, Hamiltonian simulations would have been confined to a low-energy subspace. If the variance is exponential in the moment order, it is still possible to estimate low-order moments at any energy, and this is a good stepping stone towards quantum chemistry simulations and other applications \cite{jones_chemistry_2022, dutta_simulating_2024, grimsley_adaptive_2019, jones_chemistry_2022}.

Furthermore, the maximal distance is
\begin{equation}
\begin{aligned}
R^{(\mathrm{het})}&=2|\underset{\alpha < \Omega}{\operatorname{max}} \; K(\alpha)| \, ,
\end{aligned}
\end{equation}

\noindent where $\Omega$ is the phase space truncation.  In most cases, we pick $R=2K(\Omega)$ for $\Omega$ the chosen or available phase space truncation limit. The two-mode variance generalizes easily. The full derivation is in the Supplementary Information.

\subsubsection{Homodyne measurements}
\label{sec:complexity_homodyne}
The variance of the samples $Z_i$ obtained from MGF snapshots using homodyne measurements is bounded by

\begin{equation}
\begin{aligned}
    &\operatorname{Var}^{(\mathrm{hom})}[Z] \leq\\ &\frac{(k!)^2((k+l)!)^2}{(2k+l)!}\sum^D_{n=1}\rho_{n,n} \,_3F_2\left(\frac{1}{2},-n, -(2k+l);1,1;4\right)  \, ,
\end{aligned}
\end{equation}

\noindent where we use $_pF_q$ to denote a generalized hypergeometric function. The leading order of the hypergeometric function  as $n \rightarrow \infty$ is $\mathcal{O}(n^{2k+l})$. If we set $\rho_{D, D}=1$ and the rest of the density matrix to $0$, we get that the upper bound of $\operatorname{Var}^{(\mathrm{hom})}[Z]$ has asymptotic scaling $\mathcal{O}(D^{2k+l})$, similar to the heterodyne case. The maximal distance $R$ of the homodyne estimators is

\begin{equation}
\begin{aligned}
    R^{(\mathrm{hom})} &= 2\left| \mathcal{C}_{k, l} H_{2k+l}(\Omega) \right| ,\\
     \Omega &\gg \sqrt{2(2k+l)+1} \equiv R^*\, .
\end{aligned}
\end{equation}

 The Hermite polynomials $H_n(x)$ are monotonically increasing after their last zero, $x_N < \sqrt{2n+1}$, and setting the phase space truncation $\Omega$ larger than $R^*$ guarantees an appropriate maximal distance. Both the variance and the maximal distance of a multimode system are generalized in the Supplementary Information.
 
 Finally, it is important to note that although the variance grows exponentially in the moment order, the expectation value of moment operators usually grows exponentially in the moment order itself; $\langle a^{\dagger n}a^{n}\rangle$ for instance will grow as $n$ grows. Therefore, since the expected value $\mathbb{E}[Z]$ grows itself with a higher moment order, the precision $\epsilon$ with which we want to estimate the moment (see the denominator of  Eq. \eqref{eq:measurement_number}) will be adjusted upwards as well, leading to a reduction in the number of required measurements. Derivations of the bounds as well as examples of such normalized variances $\text{Var[Z]}/\mathbb{E}[Z]$ for the photon-subtracted and entangled cat states are given in the Supplementary Information.

\section{Demonstration}

We now demonstrate the capabilities of shadow tomography on moment-generating functions using two familiar examples, namely entanglement and photon loss detection.
\label{sec:results}

\subsection{Bipartite entanglement detection}
Several CV entanglement criteria have been proposed over the years, including the Positive Partial Transpose criterion, proposed by Simon for CV systems \cite{simon_peres-horodecki_2000}. This criterion relies on the second-order moments of the state to detect entanglement. Concurrently, Duan \cite{duan_inseparability_2000} proposed a similar criterion based on second-order moments. However, second-order moments are only sufficient for detecting entanglement of Gaussian states. To extend these criteria beyond Gaussian states, Shchukin and Vogel (SV) \cite{shchukin_inseparability_2005} proposed an inseparability criterion for bipartite states which includes an infinite number of higher-order moments. Non-Gaussian states, which cannot be fully described by their first- and second-order moments, often require such higher-order inseparability criteria to be detected \cite{walschaers_non-gaussian_2021, gessner_entanglement_2017, straeter_detecting_2026}. The SV criterion can thus detect entanglement in non-Gaussian states, where lower-order criteria are insufficient.\newline
The SV criterion rests on the Negative Partial Transpose (NPT) criterion, which is a sufficient condition for two subsystems to be entangled. For operators $\hat{a}^\dagger, \hat{a}$ acting on subsystem $A$ and $\hat{b}^\dagger, \hat{b}$ acting on subsystem $B$, the SV criterion is measured from the determinant of a matrix of moment operators,

\[
D =
\begin{pmatrix}
1 &
\langle \hat{a} \rangle &
\langle \hat{a}^\dagger \rangle &
\langle \hat{b}^\dagger \rangle &
\langle \hat{b} \rangle &
\cdots \\
\langle \hat{a}^\dagger \rangle &
\langle \hat{a}^\dagger \hat{a} \rangle &
\langle \hat{a}^{\dagger 2} \rangle &
\langle \hat{a}^\dagger \hat{b}^\dagger \rangle &
\langle \hat{a}^\dagger \hat{b} \rangle &
\cdots \\
\langle \hat{a} \rangle &
\langle \hat{a}^2 \rangle &
\langle \hat{a} \hat{a}^\dagger \rangle &
\langle \hat{a} \hat{b}^\dagger \rangle &
\langle \hat{a} \hat{b} \rangle &
\cdots \\
\langle \hat{b} \rangle &
\langle \hat{a} \hat{b} \rangle &
\langle \hat{a}^\dagger \hat{b} \rangle &
\langle \hat{b}^\dagger \hat{b} \rangle &
\langle \hat{b}^2 \rangle &
\cdots \\
\langle \hat{b}^\dagger \rangle &
\langle \hat{a} \hat{b}^\dagger \rangle &
\langle \hat{a}^\dagger \hat{b}^\dagger \rangle &
\langle \hat{b}^{\dagger 2} \rangle &
\langle \hat{b} \hat{b}^\dagger \rangle &
\cdots \\
\vdots &
\vdots &
\vdots &
\vdots &
\vdots &
\ddots
\end{pmatrix} \, .
\]

A bipartite quantum system has a nonnegative partial transposition if and only if all its principal submatrices (hereafter termed \textit{SV submatrices}) have nonnegative determinants. Thus, if for a finite index set $S$ there exists a principal submatrix $(D)_S$ for which the \textit{SV determinant} $\text{det}(D)_S < 0$, the system is NPT. Finding a single negative SV determinant is enough to show that modes $A$ and $B$ are entangled. Several experimental proposals or realizations use the SV criterion to confirm entanglement in quantum systems, which has in turn helped detect cat states in optomechanical systems \cite{kanari-naish_two-mode_2022} or nonclassicality in Bose-Einstein condensates \cite{finke_observation_2016}. Beyond the SV criterion, several entanglement detection methods based on higher-order moments have been developed for non-Gaussian states \cite{miki_non-gaussian_2022, kogias_hierarchy_2015}.

Hereafter, we determine whether the SV criterion is violated for a squeezed Gaussian as well as for two families of non-Gaussian states, namely cat states and photon-subtracted states. In each case, we choose an adequate SV submatrix that shows a violation of the SV criterion analytically. Moreover, we will confirm that low-order moments show the absence of entanglement in a coherent product state. The numerical results for SV violations are plotted for both heterodyne and homodyne randomized sampling in Fig. \ref{fig:results}. In App. \ref{sec:noise}, we separately consider how noise affects the expectation values of the moments as well as the final SV determinant.

For each entangled state, we choose an SV submatrix whose determinant is negative such as to detect entanglement in the system. It should be noted however that if the state is unknown, many submatrices would need to be tested until one is found that yields a negative determinant. This is then one advantage of randomized measurements; a few thousand measurements can be used to compute many moments and evaluate different SV submatrices and their determinants. However, care must be taken because testing multiple principal submatrices on the same set of measurements introduces a multiple-testing problem, which can increase the probability of false positives unless properly accounted for. 

We present hereafter the four family of states that we will analyze.

\subsubsection{Photon-subtracted twin-beam}
\begin{figure}[htb]
\includegraphics[width=1.0\columnwidth]{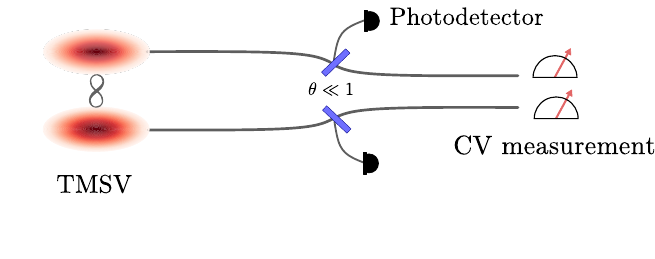}
\caption{\label{fig:photon_subtracted} A photon-subtracted twin beam is generated by first preparing a two-mode squeezed vacuum with squeezing parameter $\xi$. We couple each mode to a separate ancilla mode using a low-reflectivity beam-splitter ($\theta \ll 1$). A click on both photodetectors heralds the production of a non-Gaussian photon-subtracted state on the two main modes, after which we can make any CV measurement. }
\end{figure}

A straightforward non-Gaussian state to realize experimentally is the photon-subtracted twin-beam. Photon-subtracted states are probabilistic non-Gaussian states used in universal quantum computation on CV systems \cite{sabapathy_production_2019, su_conversion_2019, abel_simulating_2024}. Fig. \ref{fig:photon_subtracted} schematically depicts the generation of a photon-subtracted twin beam. A two-mode squeezed vacuum (TMSV) is prepared and each mode is mixed with an ancilla mode at a low-reflectivity beamsplitter. On the ancilla of each beamsplitter is placed a number-resolving photodetector. When both detectors click, we have successfully created a non-Gaussian entangled state on the main modes. The heralded state is

$$\ket{\psi_-} = \sum^\infty_{n=0} \frac{2 \tanh(r)^{(n+1)}(n+1)}{\sinh(2r)\sqrt{2\sinh(r)^2+1}} \ket{n,n} \ .$$

Owing to the strong entanglement between identical Fock states in each mode, nonzero expectation values arise only for single-mode number operators $\hat{a}^{\dagger m} \hat{a}^m$ or symmetric operators on both modes $\hat{a}^{\dagger m} \hat{a}^n \hat{b}^{\dagger m} \hat{b}^n$. One example is the number operator $\hat{n} = \hat{a}^\dagger \hat{a}$:

\begin{equation}
\begin{aligned}
\bra{\psi_-}\hat{a}^\dagger \hat{a}\ket{\psi_-} &= \sum^\infty_{n=0} \left(\frac{2 \tanh(r)^{(n+1)}(n+1)n}{\sinh(2r)\sqrt{2\sinh(r)^2+1}}\right)^2 \\
&= (3 + \text{sech}(2 r)) \sinh(r)^2 .
\end{aligned}
\end{equation}

Some low-moment-order $2\times2$ SV submatrices detect entanglement. If we consider a squeezed input state with squeezing coefficient $\xi=0.4$, we get:

\begin{equation}
\text{det}(D)_{ \{ 1, 3\}}   = 
\begin{vmatrix} \hat{a}^\dagger \hat{a} & \hat{a}^\dagger \hat{b}^\dagger \\
\hat{a} \hat{b} & \hat{b}^\dagger \hat{b}
\end{vmatrix} = -0.6605
\end{equation}

\begin{figure}[htb]
\includegraphics[width=1.0\columnwidth]{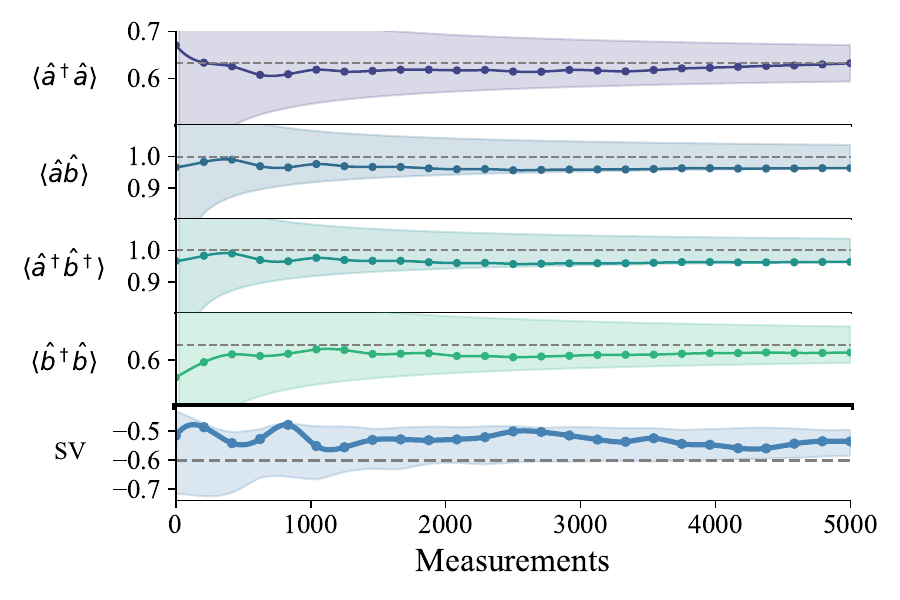}
\caption{\label{fig:moments_convergence} We display the convergence of the SV matrix components and the final SV criterion for the photon-subtracted state using homodyne measurements. The first four rows depict the expectation value of a single operator over a growing number of measurements. The light-colored bands are the standard deviations $\pm1\sigma$ derived from the exact variance in the Supplementary Information. The variance decreases similarly for all moments of $\text{det}(D)_{ \{ 1, 3\}}$, as $1/N$. The bottom row shows the evolution of a single SV value, as well as the $\pm1\sigma$ over ten independent numerical estimation rounds in the light-colored band. This figure should convey a sense of the speed at which different expectation values converge with the number of measurements. }
\end{figure}

We show how the moments and $\text{det}(D)_{ \{ 1, 3\}}$ converge to their true values in Fig. \ref{fig:moments_convergence}.

\subsubsection{Cat states}

Cat states form an important family of non-Gaussian states with interesting applications such as quantum error correction and logical qubit encoding \cite{ralph_quantum_2003, mirrahimi_dynamically_2014}. Cat states are built from superpositions of coherent states $\ket{\alpha}$. When interacting a cat state with a vacuum mode at a beam splitter, the resulting state is an entangled non-Gaussian state

\begin{equation}
\label{eq:cat_state}
\ket{\psi_\alpha} = \frac{\ket{\alpha, \alpha}-\ket{-\alpha, -\alpha}}{\sqrt{2-2e^{-|2\alpha|^2}}} \ .
\end{equation}

For the purpose of bipartite entanglement detection, we will consider the determinant of an SV submatrix

$$ \text{det}(D)_{\{ 0, 3, 6\}} = \begin{vmatrix}
1 &
\langle \hat{b}^\dagger \rangle &
\langle \hat{a}\hat{b}^\dagger \rangle\\
\langle \hat{b} \rangle &
\langle \hat{b}^\dagger \hat{b} \rangle &
\langle \hat{a}\hat{b}^\dagger \hat{b} \rangle  \\
\langle \hat{a}^\dagger \hat{b} \rangle &
\langle \hat{a}^\dagger\hat{b}^\dagger \hat{b} \rangle &
\langle \hat{a}^\dagger\hat{a}\hat{b}^\dagger \hat{b} \rangle \\
\end{vmatrix} \, .$$

Defining $\alpha'=\alpha/\sqrt{2}$, we can analytically derive the expectation values of relevant operators:
\begin{equation}
\begin{aligned}
\langle \hat{b}^\dagger \hat{b}\rangle = \langle \hat{a}^\dagger \hat{b}\rangle = \langle \hat{a} \hat{b}^\dagger \rangle &= \frac{|\alpha'|^2( 1 + e^{-|2\alpha'|^2})}{(1-e^{-|2\alpha'|^2})} \\ 
\langle \hat{a}^\dagger \hat{a}\hat{b}^\dagger \hat{b} \rangle &= |\alpha'|^4 \\
\langle \hat{b}^\dagger \rangle = \langle \hat{b} \rangle &= 0 \, .
\end{aligned}
\end{equation}

\subsubsection{Squeezed Gaussian state}
\begin{figure}[htb]
\includegraphics[width=1.0\columnwidth]{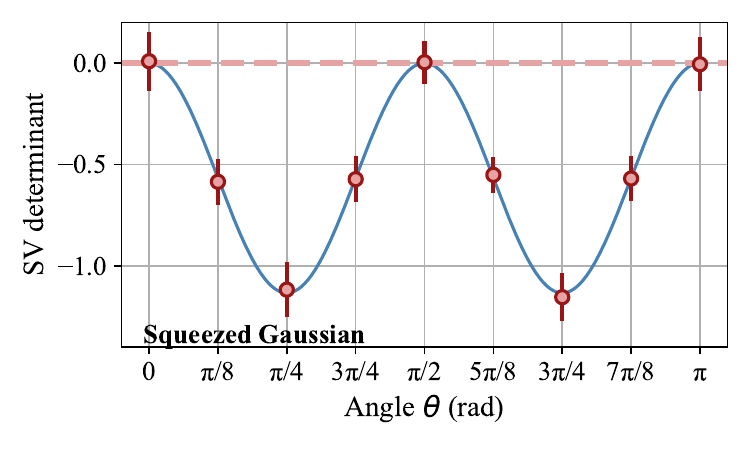}
\caption{\label{fig:squeezed_gaussian} Violation of the SV criterion for a squeezed vacuum mode coupled to a second mode with a tunable beamsplitter. The transmittivity of the beamsplitter is parametrized by $\theta$. Red dots are simulated violations for a squeezing of $\xi=1.5$ and $N=5000$ measurements. We also show the 2$\sigma$ standard deviation from 10 independent estimation rounds. We see that the 50:50 beamsplitter with $\theta=\pi/4$ or $\theta=3\pi/4$ maximises the violation, as expected. For little mixing between the two modes (for example at $\theta=0$, the system is essentially a product state, and no definitive violation is signalled.}
\end{figure}

The easiest test of entanglement is for Gaussian entangled states, which are fully described by their first- and second-order moments. More specifically, we consider a family of single-mode squeezed light, coupled to a second mode via a tunable beamsplitter. The adjustment of the transmittivity of the beamsplitter changes the amount of entanglement in the system. In Fig. \ref{fig:squeezed_gaussian}, we see how the violation of SV criterion for submatrix $\text{det}(D)_{ \{ 1, 3\}}$ is maximized for a $50:50$ beamsplitter where $\theta=\pi/4$ or $\theta=3\pi/4$. However we have violations of the SV criterion within the standard deviations for a large range of values. 

\subsubsection{Product states}
Finally, we show as an example that a specific SV submatrix which confirmed entanglement for the photon-subtracted and squeezed Gaussian states rejects the hypothesis of entanglement for a product state of two subsystems such that $\ket{\psi_{AB}} = \ket{\psi_A} \otimes \ket{\psi_B}$. The violation of the SV criterion is a sufficient condition for entanglement detection only, therefore it cannot certify separability. However, if we know that a state is separable for instance, it is a good test to verify whether some SV determinants are nonnegative as predicted. For example, for a simple submatrix $D_{ \{1, 3\} } $, we expect

\begin{equation}
\text{det} (D)_{ \{1, 3\} }  = 
\begin{vmatrix} \hat{a}^\dagger \hat{a} & \hat{a} \hat{b} \\
\hat{a}^\dagger \hat{b}^\dagger & \hat{b}^\dagger \hat{b}
\end{vmatrix} \geq 0.
\end{equation}

\begin{figure}[htb]
\includegraphics[width=1.0\columnwidth]{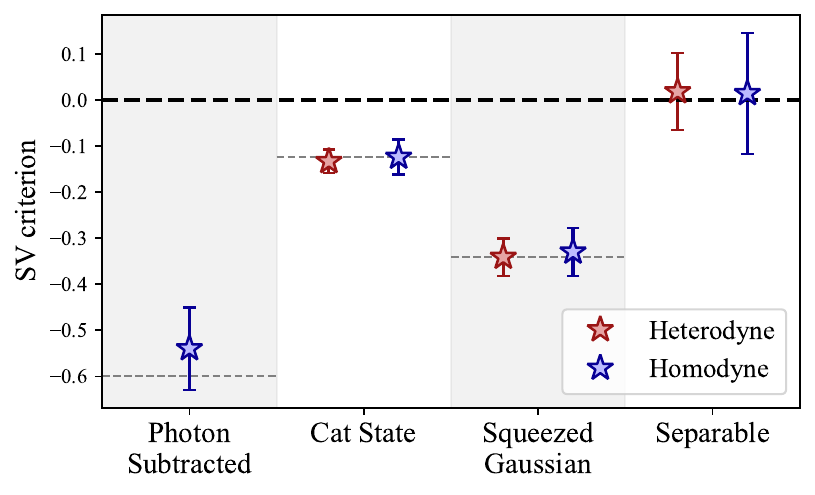}
\caption{\label{fig:results} Final results for entanglement detection using the SV criterion. We simulated ten independent runs of 5000 samples each using the CV simulation library \textit{Strawberryfields} \cite{killoran_strawberry_2019}. The library does not allow the generation of photon-subtracted states using heterodyne measurements, which explains the missing data point. The stars indicate the expectation value, the error bars are the $\pm2\sigma$ standard deviation over the ten averages, and the grey dashed line is the true value calculated analytically. We use a squeezing parameter $\xi=1.0$ and a beamsplitter angle $\theta=\pi/4$ for the squeezed Gaussian state, the entangled cat states have $\alpha=0.7$, and the TMSV used to generate the photon-subtracted state has a squeezing coefficient of $\xi=0.4$. With a Fock truncation of $D=10$, it is difficult to increase $\xi$ or $\alpha$ without loss of information. We see that the product state shows no definitive entanglement, and all other states violate the SV criterion with high probability.}
\end{figure}

Our final results in Fig. \ref{fig:results} show that the SV determinant cannot be certified negative, which is a good indication that we are dealing with a product state.

\subsection{Optical Dissipation}
Next, we use the estimation of moments from finite randomized samples to develop diagnostic tools for CV systems. For example, since states are fully characterized by their moments, the evolution of specific moments across a state preparation or evolution can give valuable information about the evolution of the CV state. Gaussian states in particular are suited for this, since up to displacement they are fully characterized by their covariance matrix $\boldsymbol{\sigma}$, which contains all second-order moments of an $M$-mode system. If we group all canonical operators into a vector $\hat{\mathbf{r}} = (\hat{q}_1,\ldots \hat{q}_m, \hat{p}_1, \ldots, \hat{p}_m)^T$ and all mode operators into another vector $\boldsymbol{\hat{\tau}} = (\hat{a}_1,\ldots \hat{a}_m, \hat{a}^\dagger_1, \ldots, \hat{a}^\dagger_m)^T $, we can change the basis with \cite{adesso_continuous_2014}

\begin{equation}
\label{eq:cov_matrix}
\begin{aligned}
\boldsymbol{\hat{\tau}} &= \boldsymbol{L}\hat{\mathbf{r}} \, ,\\
\boldsymbol{L} &= \frac{1}{\sqrt{2}} \begin{pmatrix} \boldsymbol{I} & i\boldsymbol{I} \\ \boldsymbol{I} & -i\boldsymbol{I}\end{pmatrix} \, .\\
\end{aligned}
\end{equation}

Moreover, we can write the covariance matrix in the mode operators $\boldsymbol{\tilde{\sigma}}$ in terms of the covariance matrix of the canonical operators $\boldsymbol{\sigma}$ as

\begin{equation}
\boldsymbol{\tilde{\sigma}} = \boldsymbol{L}\boldsymbol{\sigma}\boldsymbol{L}^\dagger \, .
\end{equation}

The covariance matrix $\boldsymbol{\tilde{\sigma}}$ is therefore made up of elements $\tilde{\sigma}_{ij} = \langle \hat{\tau}_i\hat{\tau}^\dagger_j + \hat{\tau}^\dagger_i\hat{\tau}_j\rangle - \langle\hat{\tau}_j\rangle\langle\hat{\tau}^\dagger_i\rangle$, for which we therefore require first and second-order moments. These moments can be estimated using our shadow tomography protocol.\newline
As mentioned in App. \ref{sec:noise}, in optical systems one of the largest sources of errors comes from optical losses. We can describe loss on a single mode $m$ using the Master equation \cite{ferraro_gaussian_2005, gaidash_lindblad_2025}:

\begin{equation}
\label{eq:master_equation}
\dot{\rho} = \frac{\Gamma_m}{2}\mathcal{L}[\hat{a}_m] \, , 
\end{equation}
\noindent with $\mathcal{L}[\hat{O}] = 2\hat{O}\rho\hat{O}^\dagger - \hat{O}^\dagger\hat{O}\rho - \rho\hat{O}^\dagger\hat{O} $ the Lindbladian associated with losses. Under this state evolution, the covariance matrix of the state evolves as 
\begin{equation}
\label{eq:covariance_evolution}
\begin{aligned}
\boldsymbol{\tilde{\sigma}}(t) &= \boldsymbol{G}^{1/2}_t\boldsymbol{\tilde{\sigma}}(0)\boldsymbol{G}^{1/2}_t + \left(\boldsymbol{I}-\boldsymbol{G}_t\right)\frac{\boldsymbol{I}}{2}\, ,\\
G_{ij} &=
\begin{cases}
e^{-\Gamma t} & \text{if } i = j = m, \\
1 & \text{if } i = j \neq m, \\
0 & \text{if } i \neq j.
\end{cases}
\end{aligned}
\end{equation}

\begin{figure}[htb]
\includegraphics[width=0.8\columnwidth]{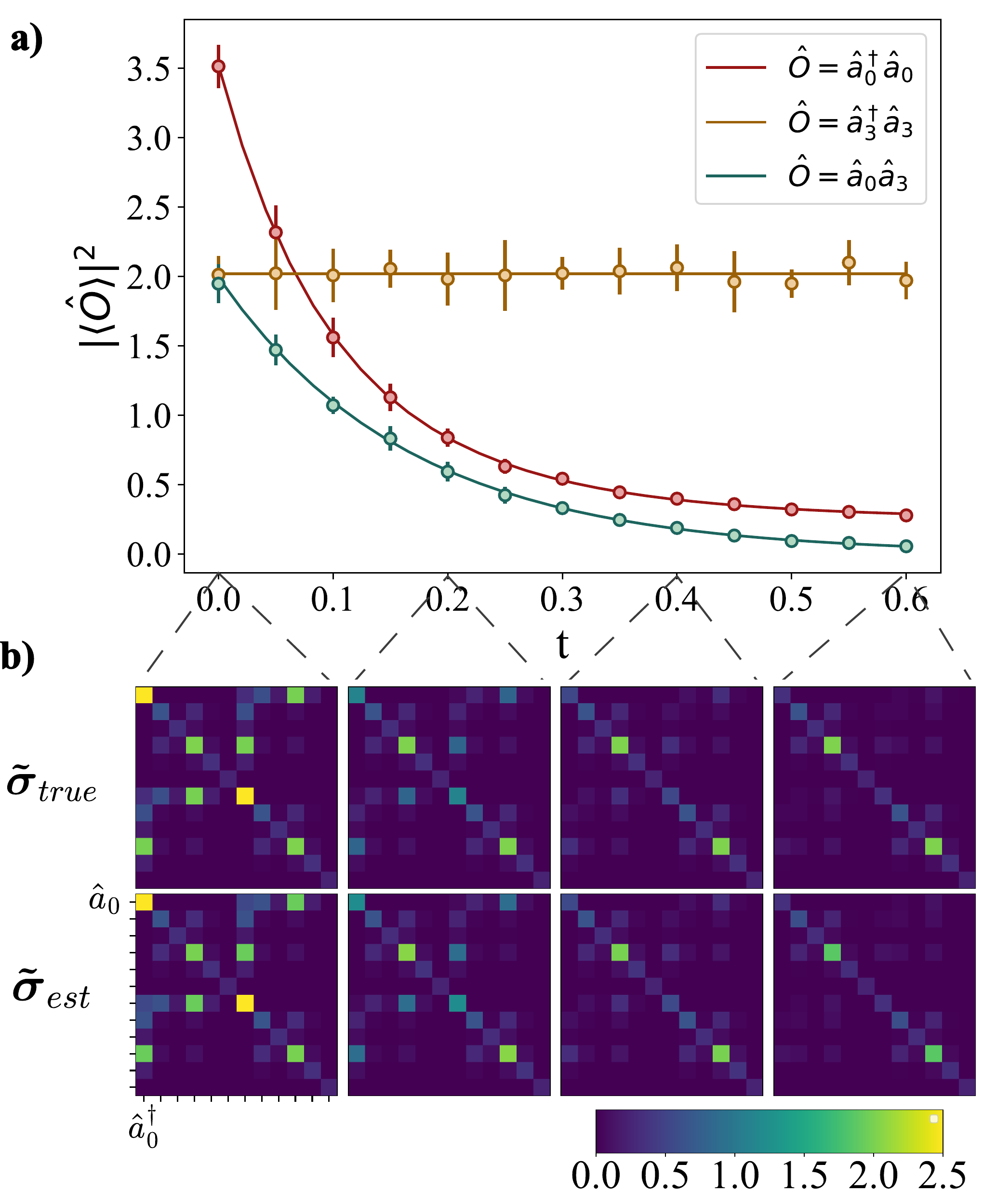}
\caption{\label{fig:photon_loss} A six-mode Gaussian state undergoes a photon loss process on the first mode. This leads to the decay of specific covariance matrix elements. In a), we show that the evolution follows an exponential decay curve in time, with a factor of either $\Gamma$ or $\frac{1}{2}\Gamma$ depending on the covariance matrix element considered. In b), we depict the full covariance matrix, expanded in $\boldsymbol{\hat{\tau}}$ for the rows and in $\boldsymbol{\hat{\tau}^\dagger}$ for the columns. We see that only terms related to $\hat{a}^\dagger_0$ or $\hat{a}_0$ decay, as expected.}
\end{figure}

We are interested in finding evidence of single-mode photon loss in the covariance matrix. We prepare some multi-mode Gaussian state, in which a squeezed light source enters a multimode interferometer,  as depicted in Fig. \ref{fig:randomized_measurements}. If we consider a lossy waveguide after the interferometer, say at mode $m=0$, the signature in the covariance matrix is clear based on Eq. \eqref{eq:covariance_evolution}. The covariance matrix elements $\tilde{\sigma}_{0, 0}, \tilde{\sigma}_{M, M}$ will decay to 0.5 as $e^{-\Gamma t}$, the matrix elements $\tilde{\sigma}_{0, M}, \tilde{\sigma}_{M, 0}$ will decay to 0 as $e^{-\Gamma t}$, and the rest of the matrix elements in rows and columns of index $\{0, M\}$ will decay to zero as $e^{-\frac{1}{2}\Gamma t}$. If these decay patterns hold, then that is proof of photon loss on the first mode. This naturally extends to any mode. Fig. \ref{fig:photon_loss} depicts the decay of these covariance elements as estimated from randomized homodyne measurements, and shows how useful shadow tomography of moments can be to identify defects in a CV system. 

\section{Conclusion}
\label{sec:conclusion}

We have studied two different types of moment-generating function shadows, derived from either randomized heterodyne or homodyne measurements on each mode. For the former, we reconstruct the well-known characteristic function of the multimode state as an average over many products of single-mode characteristic function snapshots. For homodyne measurements, we have solved the problem with the one-dimensional support of the characteristic function in phase space by developing a new MGF with full two-dimensional support in phase space. We first tested the shadow tomography protocol by estimating moments used in the Shchukin \& Vogel entanglement detection protocol. Entangled states showed strong violations  of the Shchukin \& Vogel criterion, indicating that our moments were well estimated. Secondly, we have shown how experimental teams can regularly probe their CV systems such as photonic chips to characterize the dissipation of their system with high confidence.

Our MGF shadow tomography protocol shows two considerable advantages over density matrix shadow tomography. First, the postprocessing of the samples is almost free, as our estimator kernels are simple Hermite polynomials or complex exponentials. Secondly, our methods are not limited to the characterization of low-Fock-number states. Although the variance increases exponentially with higher energy, we see no such growth when normalizing the variance with regards to the expectation value we are trying to estimate. With this fast processing of general CV moments, our randomized measurement protocol is therefore ideally suited for the simulation of bosonic Hamiltonians including for variational quantum algorithms. Our work also contributes to a better understanding and characterization of multimode CV systems, which up to now have mostly been studied for single- or double-mode systems. The next step is naturally to test this moment estimation protocol experimentally. If moments can be estimated efficiently on a given CV platform, this may open new avenues for implementing quantum algorithms based on multimode, entangled non-Gaussian states that are too complex to fully characterize using standard techniques such as full-state tomography.

\begin{acknowledgments}
Simulations were performed with the Python library Strawberryfields \cite{killoran_strawberry_2019}. M.T. and M.S. acknowledge the support
of the Singapore National Quantum Scholarship Scheme
(NQSS). L.C.K acknowledges support from the National
Research Foundation, Singapore, and the Ministry of Education, Singapore. We thank Jun Hao Hue for
his help with reviewing this manuscript.
\end{acknowledgments}

\appendix

\section{Entanglement detection with noise}
\label{sec:noise}

Beyond shot noise, where limited measurements on the system lead to an approximation error $\epsilon$, we also consider noise from imperfections in the experiment. For photonic platforms, these sources of noise are mainly from photon loss and imperfect gate noise from the phase-shifters and beamsplitters \cite{du_complete_2025}. Although CV systems routinely reach near unity fidelity on the unitary and photon losses as low as $0.1 \ \text{dB cm}^{-1}$ \cite{lenzini_integrated_2018, psiquantum_team_manufacturable_2025}, we will show how even little amounts of noise can give false negatives of the SV criterion. In this part, we therefore set out to test how shot noise and other sources of noise lead to highly variable SV determinants if the SV determinant is sensitive to the variation of the individual SV submatrix elements. This situation could arise when individual elements of the matrix are much bigger than the SV determinant itself, or when eigenvalues are very small.

To test the robustness of our randomized measurement procedure under the effects of noise, we consider the entanglement detection of the entangled cat state in Eq. \eqref{eq:cat_state} with $\alpha=1$. We chose this state because of its sensitive SV determinant. To generate the state, we mix a high amplitude cat state and a vacuum mode with a beamsplitter. We will adjust the beamsplitter transmittivity such that $T_1 = \cos(\pi/4 + \theta)$, where $\theta$ ranges from 0 to $\pi/4$; this should simulate noise on a 50:50 beamsplitter, which should have $\theta=0$. Moreover, we will consider the effects of photon loss by coupling each output mode to a separate ancilla waveguide using a beamsplitter with transmittivity $T_2$. Simulating symmetrical photon loss on both waveguides using a single parameter $T_2$ is adequate, since the reduced density matrices of each mode will be different due to the initial mixing noise. The entangled cat state at the end of the procedure is

\begin{equation}
\begin{aligned}
\ket{\psi_{noisy}} &\propto  \left|\sqrt{T_2T_1}\alpha, \sqrt{T_2(1-T_1)}\alpha\right\rangle \\
& \ \ \ \ - \left|-\sqrt{T_2T_1}\alpha, -\sqrt{T_2(1-T_1)}\alpha\right\rangle 
\end{aligned}
\end{equation}

The chosen SV determinant for the cat state is

\begin{equation}
\begin{aligned}
\text{det}(D)_{\{ 0, 3, 6\}} &= \begin{vmatrix}
1&
0 &
\langle \hat{a}\hat{b}^\dagger \rangle\\
0 &
\langle \hat{b}^\dagger \hat{b} \rangle &
0  \\
\langle \hat{a}^\dagger \hat{b} \rangle &
0 &
\langle \hat{a}^\dagger\hat{a}\hat{b}^\dagger \hat{b} \rangle \\
\end{vmatrix} ,\\ \\
\text{where} \ \langle \hat{a}^\dagger \hat{b} \rangle &= \langle \hat{a} \hat{b}^\dagger\rangle = \gamma_1\gamma_2|\alpha|^2 \eta \ ,\\
\langle \hat{b}^\dagger \hat{b} \rangle &= \gamma_2^2|\alpha|^2 \eta \ , \\
\langle \hat{a}^\dagger \hat{a} \hat{b}^\dagger \hat{b}\rangle&= \gamma_1^2\gamma_2^2 |\alpha|^4 \ .
\end{aligned}
\end{equation}

We have $\gamma_1=\sqrt{T_2T_1}, \gamma_2=\sqrt{T_2(1-T_1)}$ and the normalization factor 

\begin{equation}\eta= \frac{1-\text{exp}(-2|\gamma_1\alpha|^2-2|\gamma_2\alpha|^2)}{1+\text{exp}(-2|\gamma_1\alpha|^2-2|\gamma_2\alpha|^2)}.
\end{equation}

\begin{figure}[htb]
\includegraphics[width=1.0 \columnwidth]{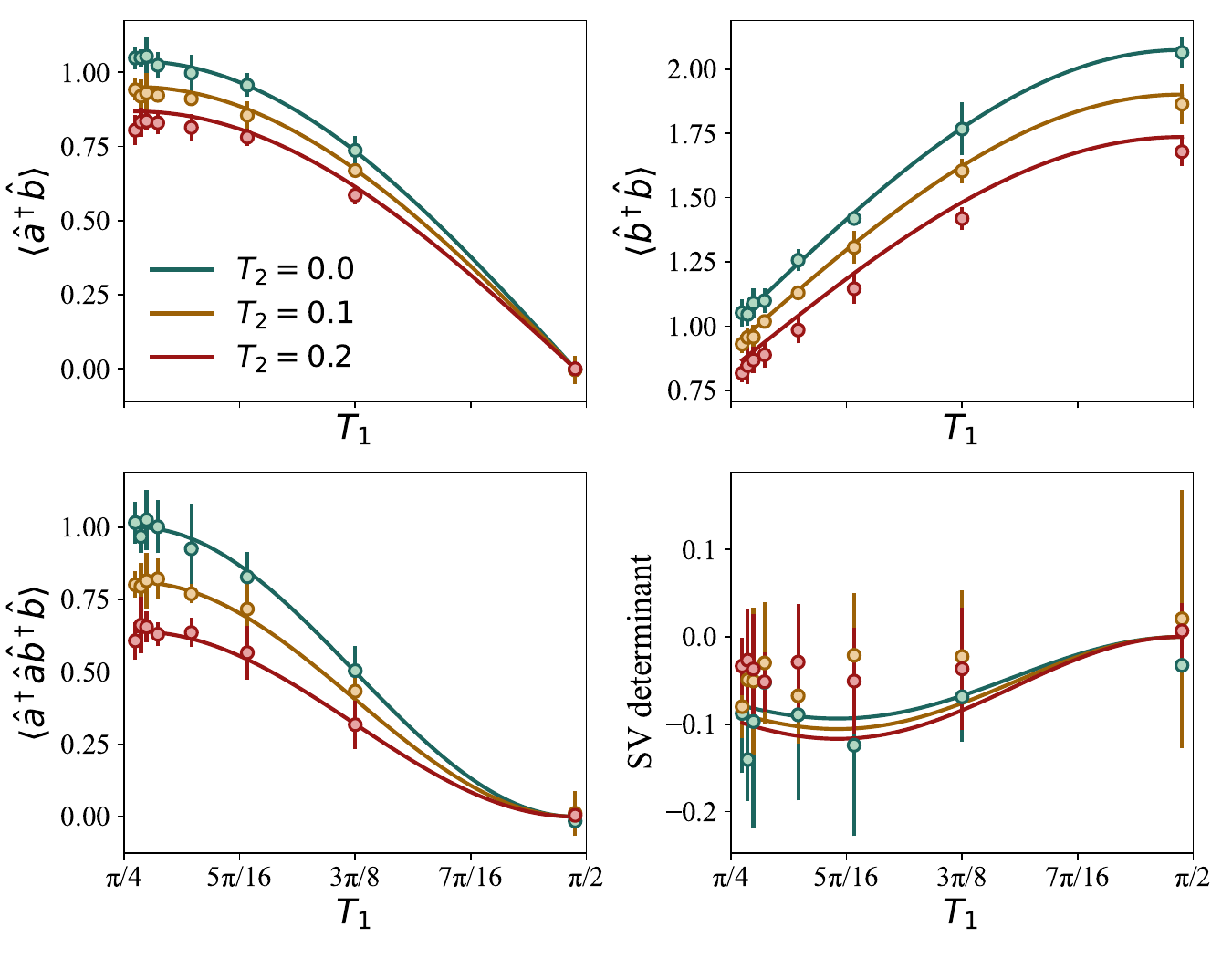}
\caption{\label{fig:noise} Effect of noise on a two-mode entangled cat state with $\alpha=1.0$. We couple each mode to an ancilla mode with a beamsplitter of transmittivity $T_2$. Furthermore, we adjust the main beamsplitter's transmittivity angle from $\pi/4$ (the 50:50 beamsplitter) to $\pi/2$. Each candle shows the mean and standard deviation over 10 independent estimation rounds of 5000 homodyne samples. The solid lines are the analytically derived expectation values. Even though all the main expectation values making up the SV criterion are very well defined on a relatively low number of samples, the SV determinant (bottom right) at this value of $\alpha$ is very sensitive to small changes in the matrix elements, and is too noisy to detect the entanglement of the cat state.}
\end{figure}

We show the effect of photon loss and beamsplitter noise on the SV submatrix and SV determinant in Fig. \ref{fig:noise}. Analytically, we expect an increase in the SV value $\text{det}(D)_{\{ 0, 3, 6\}}$ for small photon losses. Also, the initial mixing transmittivity $T_1$ plays an important role in keeping the entanglement alive, as expected. Indeed, if there were no mixing at all, we would be left with a product state. In Fig. \ref{fig:noise}, we see that individual moments can be determined to within tight bounds for different parameter regimes. However, the SV determinant for a squeezing $\xi=1$ is too sensitive to the small errors in the moments, and the SV determinant is not suitable for entanglement detection anymore.

Jacobi's formula describes the variation in the SV determinant of a matrix $\text{det}(A)$ from a perturbation in the elements of $A$ as $\partial \, \text{det}(A) / \partial A_{ij} = \text{adj}(A)_{ji}$, where $\text{adj}(A)$ is the adjugate matrix of $A$. The total variation of the determinant is then

\begin{equation}
\label{eq:delta_perturbation}
\begin{aligned}
\text{d}\left(\det(A)\right) &= \sum_{ij} \frac{\partial \, \text{det}(A)}{ \partial A_{ij} } \text{d}A_{ij} \\
&= \sum_{ij} \text{adj}(A)_{ji} \, \text{d}A_{ij} \, .
\end{aligned}
\end{equation}

If we consider the variation in A to be linearly proportional to A for small perturbations, i.e. $\text{d}A_{ij} \approx \gamma_{ij} A_{ij}$, then the SV determinant is highly impacted by both large terms in the adjugate matrix $\text{adj}(A)_{ji}$ and large matrix elements $A_{ij}$. Moreover, if $\text{det}(A) \ll A_{ij}$ for any $i, j$, then the possible impact on the SV determinant is proportionally large. Coming back to our cat state for instance, if we set the perturbation factor $\gamma_{ij}=0.1$, the weighted impact

\begin{equation}
G_{\alpha} = \sum_{ij} \left|\text{adj}(A)_{ji} \, \text{d}A_{ij} \right|
\end{equation}

\noindent gives a good estimate of the average impact of all parameter perturbations on the final SV determinant. We have $G_{0.7}$ = $0.0983$ and $G_{1.0} = 0.4385$. The SV determinants being respectively $-0.1192$ and $-0.0788$ for $\alpha=0.7$ and $\alpha=1.0$, we see how the SV determinant is particularly affected by noise in the latter case.
If the state is known beforehand, it is therefore wise to select an SV submatrix whose determinant isn't sensitive to noise on the matrix elements. If the state is unknown, Eq. \eqref{eq:delta_perturbation} can still be used to determine whether the chosen SV submatrix is adequate. Indeed, if the estimated submatrix entries produce a matrix whose determinant is highly sensitive to small perturbations, then it is better to switch to a different submatrix.


\bibliography{apssamp}
\onecolumngrid
\renewcommand{\appendixname}{}
\maketitle
\section*{Supplementary Information: Continuous-variable state moments from randomized homodyne and heterodyne measurements.}
\section*{Supp. 1: Coherent states}

Coherent states $\ket{\alpha} = e^{-\frac{|\alpha|^2}{2}} \sum^\infty_{m=0} \frac{\alpha^m}{\sqrt{m!}}\ket{m}$ display particular statistics in phase space. They are unsqueezed, minimally uncertain states whose distribution follows a Gaussian distribution. The parameter $\alpha \in \mathbb{C}$ represents a displacement in the complex phase space. Vacuum states $\ket{vac}$ are a particular type of coherent state, characterized by their null displacement. Moreover, coherent states are eigenstates of the annihilation operator, which makes it very convenient to calculate the expectation values of ladder operators analytically:

\begin{align}
\hat{a} \ket{\alpha}  &= e^{-\frac{|\alpha|^2}{2}} \sum^\infty_{m=1} \frac{\alpha^m \sqrt{m}}{\sqrt{m!}}\ket{m-1} = \alpha \ket{\alpha} \, ,\\
 \bra{\alpha} \hat{a}^\dagger  &= e^{-\frac{|\alpha|^2}{2}} \sum^\infty_{m=1} \frac{\alpha^{*m} \sqrt{m}}{\sqrt{m!}}\bra{m-1} = \bra{\alpha} \alpha^* \, .
\end{align}

A lot of derivations in the following Supplementary Information use these simple derivations to describe different bosonic states and their moments. 

\section*{Supp. 2: Wirtinger Calculus}
\label{app:wirtinger}

Treating derivatives with regards to complex variables is not as straightforward as with real variables. Instead, we must apply Wirtinger calculus \cite{koor_short_2023}. We start our treatment of complex derivatives by defining some function $f$ decomposed into its real and imaginary parts such that $f(x, y) = u(x, y) + v(x, y)$. This function is both complex itself and defined on a complex plane $z = x+iy$. If we describe a second, independent variable $z^* = x-iy$, we can now redefine $x$ and $y$ in terms of these independent variables

\begin{equation}
\begin{aligned}
x &= \frac{z + z^*}{2} \, ,\\
y &= \frac{z - z^*}{2i} \, .
\end{aligned}
\end{equation}

Since both $z$ and $z^*$ are expressed in terms of $x$ and $y$, we can apply the chain rule to find

\begin{equation}
\begin{aligned}
\frac{\partial f}{\partial z} &= \frac{\partial f}{\partial x} \frac{\partial x}{\partial z} + \frac{\partial f}{\partial y} \frac{\partial y}{\partial z}  \, ,\\
&= \frac{1}{2}\left(\frac{\partial f}{\partial x} - i\frac{\partial f}{\partial y} \right) \, ,\\
\frac{\partial f}{\partial z^*} &= \frac{\partial f}{\partial x} \frac{\partial x}{\partial z^*} + \frac{\partial f}{\partial y} \frac{\partial y}{\partial z^*} \, , \\
&= \frac{1}{2}\left(\frac{\partial f}{\partial x} + i\frac{\partial f}{\partial y} \right) \, .\\
\end{aligned}
\end{equation}

In Eq. \eqref{eq:operators_from_derivative}, we replace $z, z^*$ by $\alpha, \alpha^*$ respectively. In turn, we know that $\alpha$ is connected to the phase space coordinate system in $x, p$ by the relation $\alpha = \frac{x + ip}{\sqrt{2}}$. For a more detailed review of Wirtinger calculus, please refer to \cite{koor_short_2023}.

\section*{Supp. 3: Pattern functions}
\label{app:moment_generating_function}

If we have some operator $\hat{a}^{\dagger k+l} \hat{a}^k$, Richter \cite{richter_realistic_2000} shows that the expectation value can be derived as 
\begin{equation}
\langle \hat{a}^{\dagger k+l} \hat{a}^k\rangle = (-1)^k\frac{d^k}{d\mu^k} M_l(\mu)|_{\mu=0}
\end{equation}

from the moment-generating function 

\begin{equation}
M_l(\mu) = 
\begin{cases}
\begin{aligned}
& \text{Tr}\{\rho\hat{a}^{\dagger l}(1-\mu)^{\hat{a}^\dagger \hat{a}}\} \\
&= \sqrt{l !}\sum^{\infty}_{k=0}\sqrt{\frac{(k+l)!}{k!}} (1-\mu)^k\rho_{k, k+l}
\end{aligned}, & \text{if } l \geq 0 \, ,\\
\\
\begin{aligned}
& \text{Tr}\{\rho(1-\mu)^{\hat{a}^\dagger \hat{a}}\hat{a}^{l}\} \\
&= \sqrt{l !}\sum^{\infty}_{k=0}\sqrt{\frac{(k+l)!}{k!}} (1-\mu)^k\rho_{k+l, k}
\end{aligned}, & \text{if } l < 0 \, .
\end{cases}
\end{equation}

In Eq. \eqref{eq:M_generating_function}, we show how to evaluate this generating function from the Wigner distribution. The method relies on \textit{pattern functions} $f_{m, n}$. Pattern functions are defined as the first derivative of a product of the energy eigenfunction $u_m(x)$ and the $n$-th non-normalizable solution of the Schrödinger equation of the harmonic oscillator $v_n(x)$\cite{richter_realistic_2000},

\begin{equation}
f_{m,n} = \frac{d}{dx}u_m(x)v_n(x) \ .
\end{equation}

These pattern functions often appear when reconstructing the individual elements of a density matrix of a single mode \cite{gandhari_precision_2023}
, such that

\begin{equation}
    \begin{aligned}
    \rho_{mn} = \frac{1}{2\pi} \int^{\infty}_{-\infty} dx_\theta \int^{2\pi}_0 d\theta \, P_\theta(x_\theta)f_{m, n}(x_\theta) e^{i(m-n)\theta}.
    \end{aligned}
\end{equation}

Pattern functions being symmetric such that $f_{m, n}=f_{n, m}$, the only difference in the generating function $M_l$ between positive and negative $l$ is the phase factor $\text{exp}(i(m-n)\theta)$. It is important to note that pattern functions become increasingly difficult to evaluate at large $m, n$, and therefore any method reconstructing a state based on pattern functions is restricted to low-Fock-number states \cite{gandhari_precision_2023}. Also, the variance of the pattern functions increases with $m$, $n$.

\section*{Supp. 4: Complexity bounds}
\label{app:complexity}
\subsection*{Heterodyne Measurement}
\label{sec:complexity_heterodyne}

We have a simple formula expressing the characteristic function $\chi_s(\beta)$ in terms of the density operator $\rho$

\begin{equation}
\label{eq:chi_rho}
\chi_s(\beta) = \text{Tr}\left[\hat{D}_\beta \rho \right] e^{\frac{s}{2}|\beta|^2}
\end{equation}

The characteristic function for a given quasi-probability distribution $W_s(\alpha)$ is given by
\begin{equation}
\label{eq:characteristic_quasiprobability}
\chi_s(\beta) = \int_{\mathbb{C}} d^2\alpha \,e^{(\alpha \beta^*-\alpha^*\beta)} W_s(\alpha) 
\end{equation}

We require $\chi_1$ to determine the normally-ordered operators. This is the complex Fourier Transform of the Glauber-Sudarshan quasi-probability $P(\alpha)\equiv W_1(\alpha)$. The Glauber-Sudarshan representation expands a density operator in terms of coherent states such that $\rho = \int_{\mathbb{C}}d^2\alpha \, P(\alpha)\ket{\alpha}\bra{\alpha}$, and we get
\begin{equation}
\label{eq:glauber_sudarshan}
\chi_1(\beta) = \int_{\mathbb{C}} d^2\alpha \,e^{(\alpha \beta^*-\alpha^*\beta)} P(\alpha).
\end{equation}

Bounding the state at Fock state $|n\rangle$ is possible through the expansion of $P(\alpha)$ in the Fock basis, where

\begin{equation}
P(\alpha) = \sum_m \sum_n \bra{m}\rho\ket{n} \frac{\sqrt{m!n!}}{2\pi r(m+n)!} e^{r^2-i(m-n)\theta}\left[\left(-\frac{\partial}{\partial r}\right)^{m+n}\delta(r)\right].
\end{equation}

\noindent with $\alpha=re^{i\theta}$ in its polar form. Due to the differential operator, this form is harder to bound. If we refer again to Eq. \eqref{eq:chi_rho}, we see that $\chi_1(\beta) = \chi_{-1}(\beta)e^{|\beta|^2}$, and thus

\begin{equation}
\chi_1(\beta) = \int_\mathbb{C} d^2\alpha \, e^{(\alpha\beta^*-\alpha^*\beta)}\, Q(\alpha) \, e^{|\beta|^2}.
\end{equation}

The normally-ordered operators for a single mode are then

\begin{equation}
\begin{aligned}
\langle \hat{a}^{\dagger k+l}\hat{a}^k\rangle &= \left.\left(\frac{\partial}{\partial\beta}\right)^{k+l} \left(-\frac{\partial}{\partial\beta^*}\right)^{k}\chi_1(\beta)\right|_{\beta=0}\\
&= \int_\mathbb{C}d^2 \alpha \, K(\alpha, k, l) \,Q(\alpha),
\end{aligned}
\end{equation}

\noindent where $K_{k, l}(\alpha)$ is a polynomial in $\alpha$ acting as a kernel, weighted by the Husimi-Q distribution. The polynomial depends on the size of the moment operator, characterized by $k, l$:

\begin{center}
\begin{tabular}{@{\hspace{0.5cm}}c@{\hspace{0.5cm}}|@{\hspace{0.5cm}}c@{\hspace{0.5cm}}|@{\hspace{0.5cm}} c @{\hspace{0.5cm}}|@{\hspace{0.5cm}}c@{\hspace{0.5cm}} |@{\hspace{0.5cm}} c@{\hspace{0.5cm}}} 
 \hline
 \text{operator} & $k$ & $l$ & $K_{k, l}(\alpha)$ & $\left|K_{k, l}(\alpha)\right|^2$ \\ [0.5ex] 
 \hline\hline
 $(\hat{a}^\dagger)^0 \hat{a}^0$ & 0 & 0 & 1 & 1\\  
 \hline
 $(\hat{a}^\dagger)^1 \hat{a}^0$ & 0 & 1 & $-\alpha^*$ & $\alpha\alpha^*$\\
 \hline
  $(\hat{a}^\dagger)^0 \hat{a}^1$ &1 & -1 & $-\alpha$ & $\alpha\alpha^*$\\
 \hline
  $(\hat{a}^\dagger)^2 \hat{a}^0$ & 0 & 2 & $(\alpha^*)^2$ & $(\alpha)^2(\alpha^*)^2$ \\
  \hline
  $(\hat{a}^\dagger)^1 \hat{a}^1$ & 1 & 0 & $-1+\alpha \alpha^*$ & $1-2\alpha\alpha^* + (\alpha)^2(\alpha^*)^2 $\\
  \hline
  $(\hat{a}^\dagger)^0 \hat{a}^2$ & 2 & -2 & $\alpha^2$ & $\alpha^2(\alpha^*)^2$\\
\hline
 $(\hat{a}^\dagger)^3 \hat{a}^0$ & 0 & 3 & $-(\alpha^*)^3$ & $(\alpha)^3(\alpha^*)^3$
\end{tabular}
\end{center}

To determine the sampling efficiency, we wish to determine the precision of moment $\langle \hat{a}^{\dagger k+l}\hat{a}^k\rangle$ after $N$ measurements. To this end, it is necessary to determine the variance of the kernel $K_{k, l}(\alpha)$. The upper limit of the variance is $\text{Var[K]} \leq \mathbb{E}[|K|^2]$, where

\begin{equation}
\begin{aligned}
\mathbb{E}[|K|^2]
&= \int_\mathbb{C}d^2 \alpha \, |K_{k, l}(\alpha)|^2 \,Q(\alpha).
\end{aligned}
\end{equation}

If we calculate the integral over each term in the polynomial $|K_{k, l}(\alpha)|^2$ separately, we can derive a general form of the variance for a single monomial $\gamma_{p, q} \alpha^{*p}\alpha^q$. Expanding the Husimi-Q function into its Fock basis,

\begin{equation}
\begin{aligned}
Q(\alpha) =  \frac{1}{\pi} \bra{\alpha}\rho\ket{\alpha} &= \frac{1}{\pi} \bra{\alpha}\sum_{m, n} \rho_{m, n} | m\rangle \langle n|\alpha\rangle\\
&= \frac{1}{\pi}\sum_{m, n} e^{-|\alpha|^2}\sum_{p, q}\frac{\alpha^{*p}\alpha^q}{\sqrt{p!\,q!}} \rho_{m, n} \langle p| m\rangle \langle n|q\rangle\\
&=\frac{1}{\pi}\sum_{m, n} e^{-|\alpha|^2}\frac{\alpha^{*m}\alpha^n}{\sqrt{m!\,n!}} \rho_{m, n} \, ,
\end{aligned}
\end{equation}

\noindent we can multiply the Husimi-Q function with each term in the polynomial $|K_{k, l}(\alpha)|^2$. For a single mode and a single monomial, we therefore have
\begin{equation}
\label{eq:husimi_factorial}
\begin{aligned}
\frac{\gamma_{p,q}}{\pi} \int \text{d}\alpha \, \sum_{m, n} e^{-|\alpha|^2}\frac{\alpha^{*m+p}\alpha^{n+q}}{\sqrt{m!\,n!}} \rho_{m, n}  &= \frac{\gamma_{p,q}}{\pi} \sum_{m, n} \frac{\rho_{m, n}}{\sqrt{m!\,n!}} \int^{\infty}_{0}\int^{\pi}_{0} \text{d}r \,\text{d}\theta  \  r \cdot r^{m+p+n+q} e^{i(m+p-n-q)\theta} e^{-r^2}\\
&= \gamma_{p,q}\sum_{\substack{m,n \\ m+p=n+q}} \frac{\rho_{m, n}}{\sqrt{m!\,n!}}\int^{\infty}_{0} \text{d}r \, r^{2(m+p)+1} e^{-r^2} \\&=\gamma_{p,q}\sum_{\substack{m,n \\ m+p=n+q}} \frac{\rho_{m, n}}{\sqrt{m!\,n!}}\Gamma(m+p+1) = \gamma_{p,q}\sum_{\substack{m,n \\ m+p=n+q}} \rho_{m, n}\frac{(m+p)!}{\sqrt{m!\,n!}}
\end{aligned}
\end{equation}

The sum over $m, n$ must follow the constraint $m+p=n+q$. Evidently, we are only using terms of the density matrix along a specific diagonal. When $p=q$, we have the main diagonal of the density matrix, else we are removed by $p-q$ from the main diagonal. The main diagonal terms can have a value $|\rho_{m, m}| \leq 1$, off-diagonal terms are bounded by $|\rho_{m, n}| \leq \frac{1}{2}$ when $m\neq n$. We have no general proof for which state has the highest possible variance for our heterodyne sampling protocol, therefore we will assume all diagonal terms are 1 and all off-diagonal coefficients are $\frac{1}{2}$ to estimate an upper bound. Then, we would like to understand the complexity of each factorial fraction in the final result of \eqref{eq:husimi_factorial}, and we get

\begin{equation}
\label{eq:leading_term}
\begin{aligned}
\frac{(m+p)!}{\sqrt{m!\,n!}} = \frac{(n+q)!}{\sqrt{(n-p+q)!\,n!}} &\stackrel{\text{(i)}}{=}\frac{\Gamma(n+q+1)}{\Gamma(n-p+q+1)^{1/2}\Gamma(n+1)^{1/2}}\\
&=\log\Gamma(n+q+1) - \frac{1}{2}\Gamma(n-p+q+1)-\frac{1}{2}\Gamma(n+1)\\
&=n^{\frac{p+q}{2}}\left(1 + \mathcal{O}\left(\frac{1}{n}\right)\right) .
\end{aligned}
\end{equation}

In (i) we use the Gamma function $\Gamma(n+1)=n!$. As we can see, the leading order is $n^{(p+q)/2}$. The highest possible value of $n$ on the diagonal where $m+p=n+q$ will then be the leading order of the full sum \eqref{eq:husimi_factorial}. 

\begin{figure}[htb]
\includegraphics[width=0.8\columnwidth]{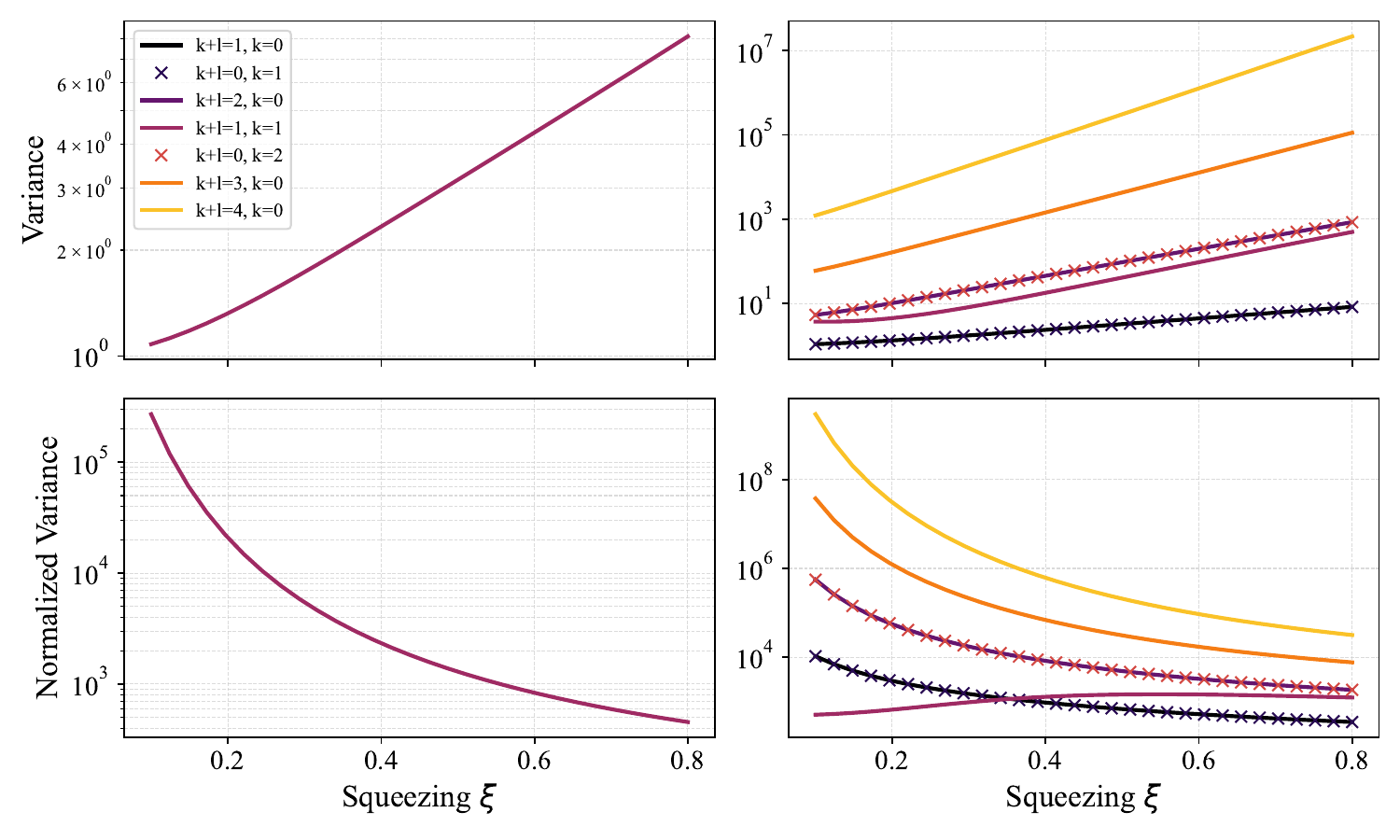}
\caption{\label{fig:tmsv_variance} Variance and normalized variance for moment operators acting on a single mode (left-hand side) and operators of the same degree acting on both modes (right-hand side) for a photon-subtracted state sampled with heterodyne measurements. Single-mode variance curves are not drawn if the expectation value is $0$. For the normalized variance, we adjusted the $\epsilon$ of the Bernstein inequality to 5\% of the expectation value. We note that although the variance increases with the squeezing, the normalized variance is decreasing, meaning highly-squeezed states need less measurements to get the estimator within a certain percentage of its final value.}
\end{figure}

\begin{figure}[htb]
\includegraphics[width=0.8\columnwidth]{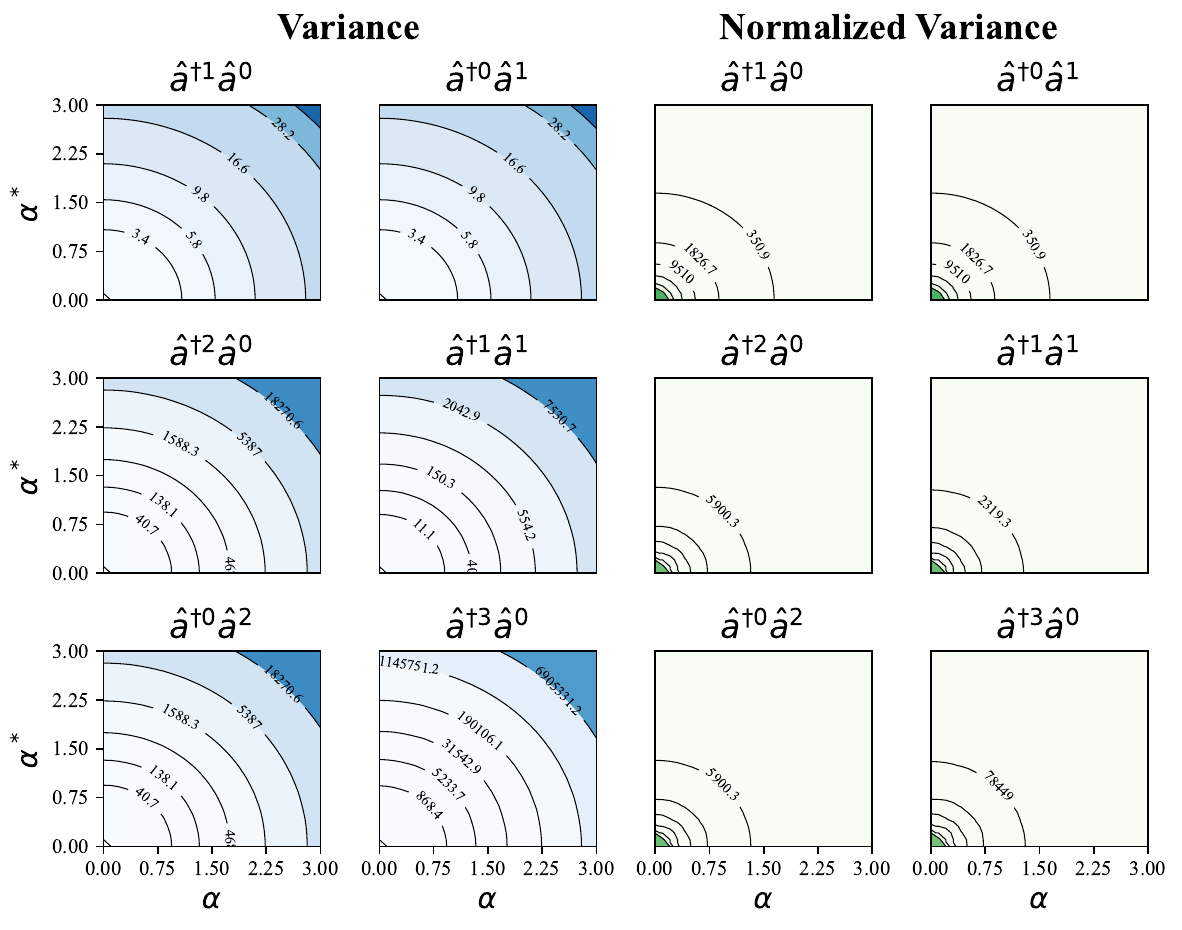}
\caption{\label{fig:cat_variance_heterodyne} Entangled cat state moment variance (in blue) and normalized variance (in green) from heterodyne measurements, \textit{applied to both modes identically}. The darker the color, the larger the variance, and the more measurements will be required to get the expectation within $\epsilon$ of the true value. For the normalized variance, we divide the variance by $\epsilon^2=\left(0.05\langle(\hat{a}^\dagger)^{k+l} \hat{a}^k\rangle\right)^2$, as done in Eq. \eqref{eq:measurement_number}. Interestingly, the normalized variance decreases with a higher $\alpha$ value.}
\end{figure}

For two modes, we have a sum over terms in the form of 

\begin{equation}
\frac{(m_1+p_1)!}{\sqrt{m_1! n_1!}} \frac{(m_2+p_2)!}{\sqrt{m_2! n_2!}} \, ,
\end{equation}

\noindent and applying the same approximation procedure as in Eq. \eqref{eq:leading_term} we see that the leading term of the two-mode variance is just the product of the leading terms of each individual mode.
In Fig. \ref{fig:tmsv_variance} we plot the variance and normalized variance for the photon-subtracted state. In Fig. \ref{fig:cat_variance_heterodyne}, we plot the same variance profiles for a family of cat states parametrized by the coherent state displacements $\ket{\alpha}$, where $0\leq\text{Re}\, \alpha \leq 3.0, 0 \leq \text{Im}\,\alpha \leq 3.0$.

\subsection*{Homodyne Measurement}
\label{sec:app_complexity_homodyne}

For the homodyne setting, if we are considering bounds on an expectation value $\mathbb{E}[Z]= \langle \hat{a}^{\dagger k+l} \hat{a}^{k} \rangle$, we need to determine the bounds on the single-shot estimator. The single-shot estimator is sampled from the Radon-transformed Wigner distribution $P_\theta(x_\theta)$ (see Eq. \eqref{eq:moment_from_hermite}) and is 

\begin{equation}
    \label{eq:complexity_homodyne}
    Z_j(k, l) = \mathcal{C}_{k,l} \times H_{2k+l}(x_\theta^{(j)})e^{-i l \theta_j},
\end{equation}

\noindent where $\mathcal{C}_{k,l} = \left[\pi\sqrt{2^{2k+l}} \binom{2k+l}{k+l}\right]^{-1}$. 

Bernstein's inequality \cite{rebeschini_bernsteins_nodate} gives a bound on the probability that some randomly sampled variable is within $\epsilon$ of its expected value $Z$:

\begin{equation}
    \label{eq:bernstein}
    \mathbb{P}\left( \left|\frac{1}{N} \sum_{j=1}^N Z_j - \mathbb{E}[Z]  \right|\geq \epsilon \right) \leq 2\,\text{exp}\left( -\frac{N \epsilon^2/2}{\operatorname{Var}[Z] + R \epsilon /3}\right)
\end{equation}

\noindent with R the maximal distance $|Z_i - \mathbb{E}[Z]|$ and N the number of samples. Since we do not make any assumptions about the nature of the distribution of $Z$, we take the variance to be bounded by the maximum possible variance on a range $ x_\theta \in [-\Omega_{max}, \Omega_{max}]$. A first estimate of the variance yields

\begin{equation}
    \label{eq:variance}
    \begin{aligned}
    \operatorname{Var}[Z] &= \mathbb{E}[|Z|^2] - |\mathbb{E}[Z]|^2 \\
    &= \mathbb{E}\left[\left|\mathcal{C}_{k,l} \, H_{2k+l}(\Omega_{max}) e^{-i l \theta}\right|^2\right] - \left|\mathbb{E}\left[\mathcal{C}_{k,l} \, H_{2k+l}(\Omega_{max}) e^{-i l \theta}\right]\right|^2\\
    &\leq |\mathcal{C}_{k,l}|^2 \mathbb{E}\left[\left|H_{2k+l}(\Omega_{max})\right|^2\right]\\
    &= |\mathcal{C}_{k,l}|^2 \left|H_{2k+l}(\Omega_{max})\right|^2 \, ,
    \end{aligned}
\end{equation}

\noindent as long as $\Omega_{\mathrm{max}} \gg \sqrt{2(2k+l)+1/2}$, the last zero of the Hermite polynomial. This bound has high polynomial growth in the truncation $\Omega$, which gives a rather poor upper bound. Alternatively, we can review the moment-generating function. Instead of placing bounds $\Omega_{max}$ on the phase space, we can place bound $D$ on the Fock space. This is experimentally also a more tractable approach, as photon statistics are more prevalent and therefore it is easy to determine the highest occupation number. 
Looking back at Eq. \eqref{eq:variance}, our estimator in this case is $Z^{(M)}=\prod^M_{j=1} \mathcal{C}_{k_j, l_j}H_{2k_j+l_j}(x)e^{-il\theta}$ for some moment $\braket{\hat{a}^{\dagger  k_1+l_1} \hat{a}^{k_1} \dots\hat{m}^{\dagger  k_m+l_m} \hat{m}^{k_m}}$ on a system of $M$ modes. We can devise an upper bound with the first term $\mathbb{E}[|Z|^2]$ only, disregarding the square of the expectation value. In the single mode case where $M=1$, we then get

\begin{equation}
\label{eq:variance_onemode}
\begin{aligned}
\mathbb{E}[(Z^{(1)})^2] &= \frac{1}{\pi}\int^{\pi}_0d\theta \int_{-\infty}^\infty dx_\theta \,P_\theta(x_\theta) \left| \pi \mathcal{C}_{k, l} H_{2k+l}(x_\theta)e^{-il\theta}\right|^2 \\  
&= \frac{1}{\pi}\int^{\pi}_0d\theta \int_{-\infty}^\infty \, dx_\theta \sum^{D}_{m, n=0} \rho_{m, n} \frac{e^{-i(m-n)\theta}}{(\pi)^{1/2}} \text{exp}\{-x_\theta^2\} \frac{H_n(x_\theta)H_m(x_\theta)}{\sqrt{2^{n+m}n!m!}}\pi^2\mathcal{C}_{k, l}^2 H_{2k+l}(x_\theta)^2 \\
&=  \int_{-\infty}^\infty dx_\theta \sum^{D}_{m=0} \rho_{m, m} \frac{1}{(\pi)^{1/2}} \text{exp}\{-x_\theta^2\} \frac{1}{\sqrt{2^{2m}(m!)^2}}\, \pi^2\mathcal{C}_{k, l}^2 \, H_m(x_\theta)^2 \,H_{2k+l}(x_\theta)^2.
\end{aligned} 
\end{equation}

\noindent where we have expressed the Radon-transformed Wigner distribution in terms of the density matrix $\rho$ with the relationship given in \cite{richter_determination_1999}. 
 Now, we will solve the integral to extract the variance of the state. We use the product relation and the orthogonality relation for Hermite polynomials \cite{feldheim_hermite, watson_polynomials_hermite, gradstejn_integrals},

\begin{equation}
\begin{aligned}
H_m (x) H_n(x) &= 2^{n} n! \sum^{n}_{r=0}\binom{m}{n-r}\frac{1}{2^r r!}\, H_{m - n + 2r}(x),\\
\frac{1}{\sqrt{\pi}}\int \text{d}x e^{-x^2}\ H_m(x) H_n(x)  &= 2^m m! \, \delta_{m, n} \, . 
\end{aligned}
\label{eq:hermite_identities}
\end{equation}
An integral over three Hermite polynomials then gives us
\begin{equation}
\frac{1}{\sqrt{\pi}}\int^{+\infty}_{-\infty} \mathrm{d}x \, e^{-x^2} H_k(x) H_m(x) H_n(x) = 2^{n}n! \sum^{n}_{j=0}\binom{m}{n-j} \frac{2^k k!}{2^j j!} \ \delta_{k, m-n+2j} .
\end{equation}
To solve the four Hermite integral in Eq. \eqref{eq:variance_onemode}, we first decompose one of the square Hermite polynomials into a sum over evenly-spaced Hermite polynomials according to \eqref{eq:hermite_identities}. Then, we apply the triple integral relation, and we get

\begin{equation}
\label{eq:homodyne_scaling}
\begin{aligned}
\mathbb{E}[(Z^{(1)})^2] &= \pi^2\mathcal{C}^2_{k,l} \sum^D_{m=0} \rho_{m,m} \sum^m_{r=0}\binom{m}{m-r} \frac{1}{2^r r!} \frac{1}{\sqrt{\pi}} \int^{+\infty}_{-\infty} \text{d}x \, e^{-x^2} H_{2r}(x) H_{2k+l}(x) \\
&=\pi^2\mathcal{C}^2_{k,l} \sum^D_{m=0} \rho_{m,m} \sum^m_{r=0}\binom{m}{m-r} \frac{1}{2^r r!} 2^{2k+l}(2k+l)! \sum^{2k+l}_{j=0}\binom{2k+l}{2k+l-j} \frac{1}{2^jj!} 2^{2r} (2r)! \ \delta_{r, j} \\
&= \sum^D_{m=0} \rho_{m, m} \sum^{2k+l}_{r=0}\binom{m}{m-r}\frac{(k+l)! (k+l)! k! k!}{(2k+l-r)!r!}\frac{(2r)!}{(r!)^2} \\
&= \frac{(k!)^2((k+l)!)^2}{(2k+l)!} \sum^D_{m=0} \rho_{m, m} \ _3F_2\left(\frac{1}{2},-m,-(2k+l);1,1;4\right)\\
&\leq  \frac{(k!)^2((k+l)!)^2}{(2k+l)!}  \ _3F_2\left(\frac{1}{2},-D,-(2k+l);1,1;4\right) \ \ .
\end{aligned}
\end{equation}

\noindent The final result is expressed in terms of a generalized hypergeometric function (GHF) $f_Q(D) \equiv \, _3F_2\left(\frac{1}{2},-D,-(2k+l);1,1;4\right)$, with $Q=2k+l$ the moment order. The final inequality is valid because $f_Q(D) \geq f_Q(D-1)$ if $D\geq Q, D\geq1$. Indeed, if we expand the GHF as a sum,

\begin{equation}
f_Q(D) = \sum^{\mathrm{min}(Q, D)}_{n=0}\frac{\left(\frac{1}{2}\right)_n\left(-Q\right)_n\left(-D\right)_n 4^n}{\left(n!\right)^3} \, ,
\end{equation}

\noindent where $\left(a\right)_n = a(a+1)\dots(a+n-1)$ is a Pochhammer-like symbol, we can use this expression to comment on the difference $\Delta f_Q \equiv f_Q(D)-f_Q(D-1)$. This is

\begin{equation}
\Delta f_Q = \sum^{\mathrm{min}(Q, D)}_{n=0}\frac{\left(\frac{1}{2}\right)_n\left(-Q\right)_n 4^n}{\left(n!\right)^3}\left(\left(-D\right)_n - \left(-D+1\right)_n\right) \, .
\end{equation}

The difference between two Pochhammer terms is 

\begin{equation}
\left(\left(-D\right)_n - \left(-D+1\right)_n\right) = -n \left(-D+1\right)_{n-1} \, .
\end{equation}

The total difference $\Delta f_Q$ is finally simplified to 

\begin{equation}
\begin{aligned}
\Delta f_Q &= \sum^{\mathrm{min}(Q, D)}_{n=1}\frac{\left(\frac{1}{2}\right)_n\left(-Q\right)_n 4^n}{\left(n!\right)^3}(-1)n\left(-D+1\right)_n \, \\
&=\sum^{\mathrm{min}(Q, D)}_{n=1}\frac{\left(\frac{1}{2}\right)_n\left(Q\right)_n 4^n}{\left(n!\right)^2(n-1)!}\left(D-1\right)_n \geq 0 \ \ \text{if} \ \ D\geq1.
\end{aligned}
\end{equation}

This concludes the proof of the behaviour of different GHF $f_q(D)$. 

The multimode variance estimate follows a similar procedure. For example, if we consider two modes, a lengthy but straightforward extension of the work in \cite{wunsche_tomographic_1996} gives us the moment

\begin{equation}
\langle\hat{a}^{\dagger k_1+l_1}\hat{a}^{k_1}\hat{b}^{\dagger k_2+l_2}\hat{b}^{k_2}\rangle = \mathcal{C}_{k_1, l_1} \mathcal{C}_{k_2, l_2}\iint^\pi_0 \mathrm{d}\theta_1\mathrm{d}\theta_2 e^{-i (l_1 \theta_1 + l_2 \theta_2)} \iint^\infty_{-\infty} \mathrm{d}x_1 \mathrm{d}x_2 \, P_{\theta_1, \theta_2}(x_1,x_2) \,H_{2k_1+l_1}\left(x_1\right)H_{2k_2+l_2}\left(x_2\right) \, .
\end{equation}

The variance is then

\begin{equation}
\label{eq:variance_twomode}
\begin{aligned}
\mathbb{E}[(Z^{(2)})^2] &= \pi^4 \int^\infty_{-\infty}dx_1dx_2 \left(\sum^D_{m, \mu=0}\rho_{m,m,\mu,\mu}\frac{1}{\pi} \text{exp}\{-x_1^2 -x_2^2\}\frac{1}{\sqrt{2^{2(m+\mu)}(m!)^2(\mu!)^2}}\right)\\& \hspace{6cm} \times \mathcal{C}_{k_1, l_1}^2\mathcal{C}_{k_2, l_2}^2H_{m}(x_1)^2H_{\mu}(x_2)^2H_{2k_1+l_1}(x_1)^2 H_{2k_2+l_2}(x_2)^2 \\  
&= \frac{(k_1!)^2((k_1+l_1)!)^2}{(2k_1+l_1)!} \frac{(k_2!)^2((k_2+l_2)!)^2}{(2k_2+l_2)!}  \\
& \hspace{2cm}\times \sum^D_{m=0, \, \mu=0} \rho_{m, m, \mu, \mu} \, _3F_2\left(\frac{1}{2},-m,-(2k_1+l_1);1,1;4\right) \, _3F_2\left(\frac{1}{2},-\mu,-(2k_2+l_2);1,1;4\right)\\
&\leq \frac{(k_1!)^2((k_1+l_1)!)^2}{(2k_1+l_1)!} \frac{(k_2!)^2((k_2+l_2)!)^2}{(2k_2+l_2)!} \ _3F_2\left(\frac{1}{2},-D, -(2k_1+l_1);1,1;3\right)\ _3F_2\left(\frac{1}{2},-D, -(2k_2+l_2);1,1;3\right).
\end{aligned}
\end{equation}

The leading order term of the hypergeometric function is $\mathcal{O}(D^{2k+l})$. Therefore, the variance for a single mode moment grows polynomially in the Fock space truncation $D$. Although the variance can reach high levels for large $D$, which can lead to unrealistically high measurement numbers, it is important to note that the expectation value does itself grow with the power of the moments. In many cases, this will suppress the growth in Eq. \eqref{eq:measurement_number} as it increases the desired accuracy $\epsilon$ in the denominator. 

\begin{figure}[htb]
\includegraphics[width=0.8\columnwidth]{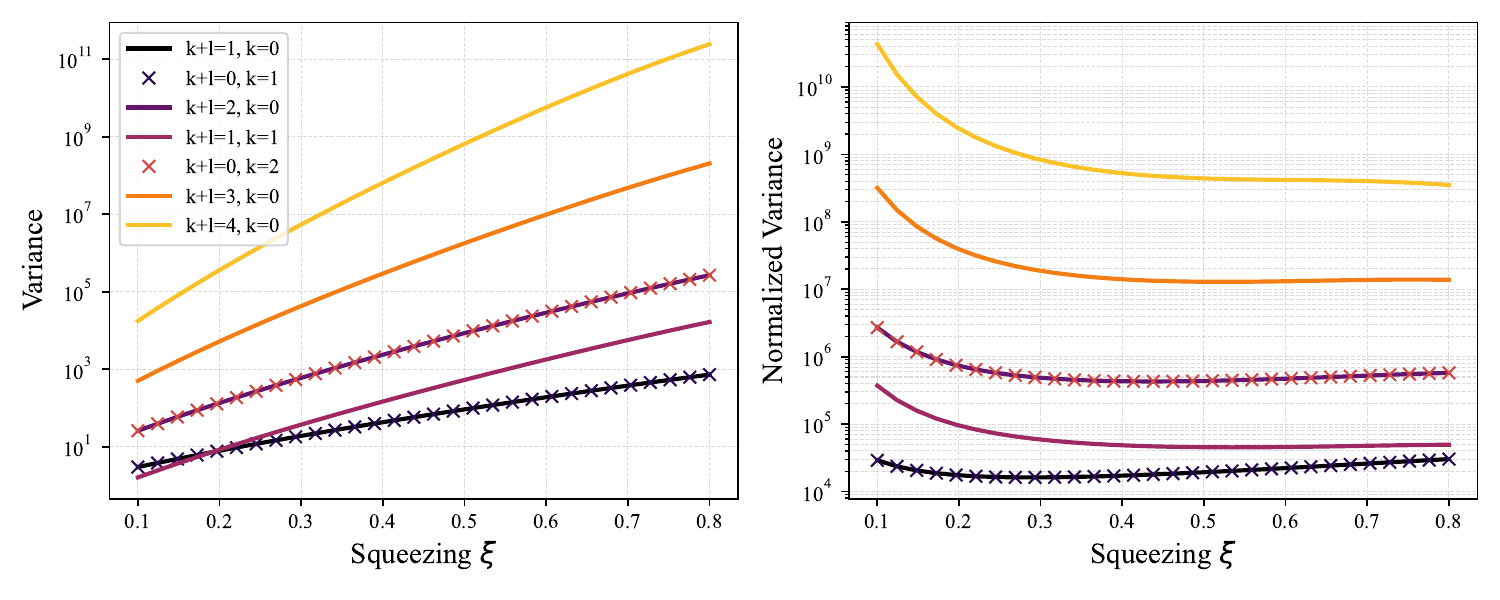}
\caption{\label{fig:tmsv_variance_homodyne} Variance and normalized variance of moment operators \textit{acting on both modes identically} for a photon-subtracted state sampled with homodyne measurements. For the normalized variance, we adjusted the $\epsilon$ of the Bernstein inequality to 5\% of the expectation value. We note that although the variance increases with the squeezing, the normalized variance is decreasing, meaning highly-squeezed states need less measurements to get the estimator within a certain percentage of its final value.}
\end{figure}

\begin{figure}[htb]
\includegraphics[width=0.8\columnwidth]{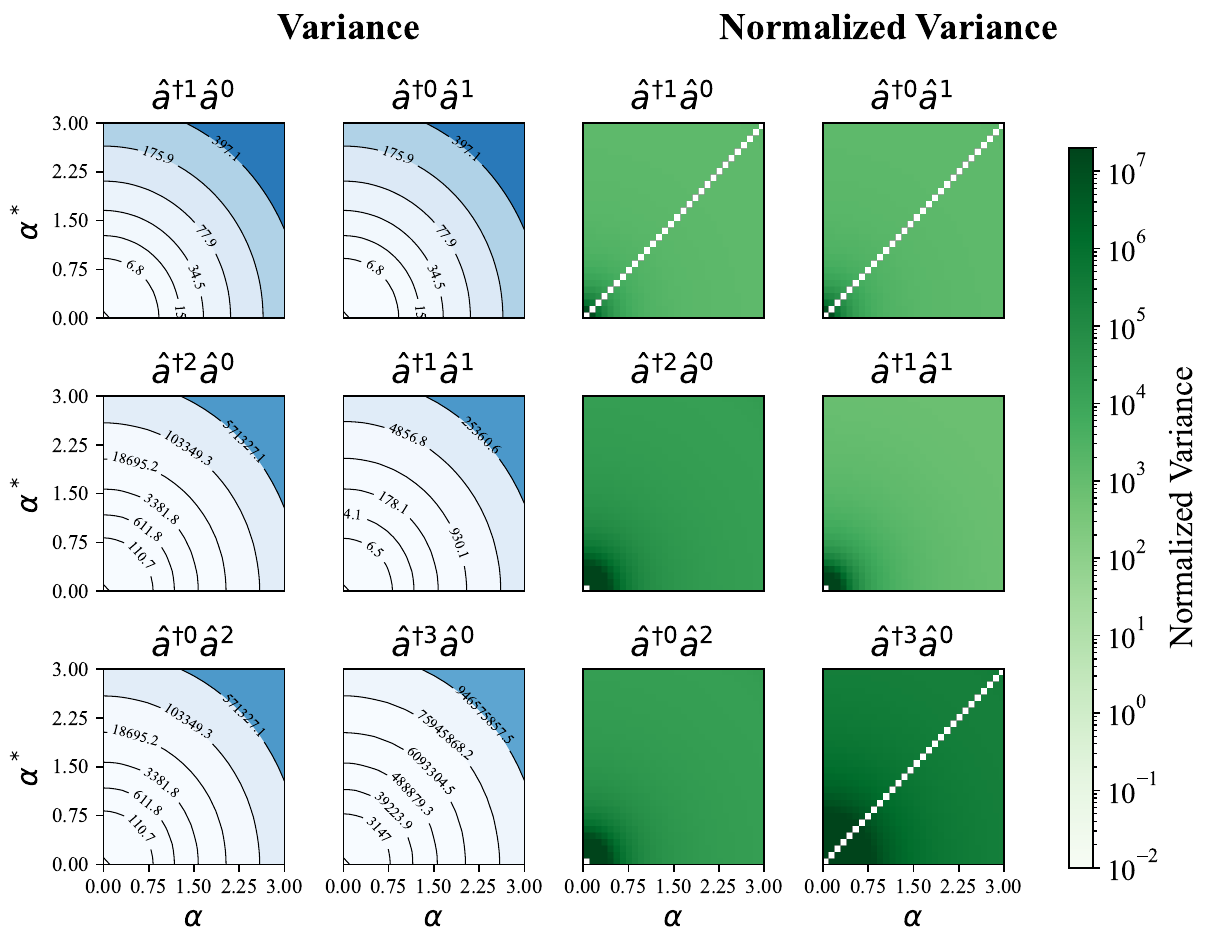}
\caption{\label{fig:cat_state_variance_homodyne} Entangled cat state moment variance (in blue) and normalized variance (in green) from homodyne measurements, \textit{applied to both modes identically}. The darker the color, the larger the variance, and the more measurements will be required to get the expectation within $\epsilon$ of the true value. For the normalized variance, we divide the variance by $\epsilon^2=\left(0.05\langle(\hat{a}^\dagger)^{k+l} \hat{a}^k (\hat{b}^\dagger)^{k+l} \hat{b}^k\rangle\right)^2$, as done in Eq. \eqref{eq:measurement_number}. Interestingly, the normalized variance decreases with a higher $\alpha$ value.}
\end{figure}

In Fig. \ref{fig:tmsv_variance_homodyne}, we show the scaling for the photon-subtracted state, as a function of the squeezing parameter $\xi$. We see that the variance is higher than that calculated for the heterodyne measurement protocol. In Fig. \ref{fig:cat_state_variance_homodyne}, we also plot the variance for several moments of cat states parameterized by their coherent state displacement $\alpha$.

\section*{Supp. 5: Common States}\label{app:states}

\subsection*{Squeezed Gaussian State}

Gaussian states are easy to represent classically, since all of their information is contained within the first and second moments. These moments are denoted by $\mu \in \mathbb{R}^{2m}$ and $V \in \mathbb{R}^{2m \times 2m}$ for $m$ modes. We start with a vacuum state, which we squeeze at an angle $\theta$ with squeezing coefficient $\xi$. We then pass the state into a beamsplitter with mixing angle $\phi$. The resulting state retains its Gaussian character. Detecting entanglement is trivial using the determinant of a small SV submatrix:

\begin{equation}
    \label{eq:squeezed_coherent_sv}
    \text{det}(D)_{ \{ 1, 3\}} = \begin{vmatrix}
    \hat{a}^\dagger\hat{a} & \hat{a}^\dagger\hat{b}^\dagger \\
    \hat{a}\hat{b} & \hat{b}^\dagger \hat{b}
    \end{vmatrix}.
\end{equation}

We can derive the SV determinant analytically for a state by using known relations between the ladder operators and the $x-p$ operators,
\begin{equation}
    \begin{aligned}
    \hat{a}^\dagger = \frac{1}{\sqrt{2}}(\hat{x}_a - i\hat{p}_a), & \hspace{1cm} \hat{a}= \frac{1}{\sqrt{2}}(\hat{x}_a + i\hat{p}_a), \\
    \hat{b}^\dagger = \frac{1}{\sqrt{2}}(\hat{x}_b - i\hat{p}_b), & \hspace{1cm} \hat{b}= \frac{1}{\sqrt{2}}(\hat{x}_b + i\hat{p}_b).\\
    \end{aligned}
\end{equation}

The required expectation values are
\begin{equation}
    \begin{aligned}
    \langle \hat{a}^\dagger \hat{a} \rangle  &= \frac{1}{2} (\hat{x}_a^2 + \hat{p}_a^2 - 1) = \frac{1}{2} (V_{00}^2 + V_{22}^2 - 1), \\
    \langle \hat{a}\hat{b} \rangle  &= \frac{1}{2} (\hat{x}_a \hat{x}_b + i\hat{x}_a\hat{p}_b + i \hat{x}_b\hat{p}_a - \hat{p}_a \hat{p}_b) = \frac{1}{2} (V_{01} + iV_{03} + i V_{12} - V_{23}),\\
    \langle \hat{a}^\dagger \hat{b}^\dagger \rangle  &= \frac{1}{2} (\hat{x}_a \hat{x}_b - i\hat{x}_a\hat{p}_b - i \hat{x}_b\hat{p}_a - \hat{p}_a \hat{p}_b) = \frac{1}{2} (V_{01} - iV_{03} - i V_{12} - V_{23}),\\ 
    \langle \hat{b}^\dagger \hat{b} \rangle  &= \frac{1}{2} (\hat{x}_b^2 + \hat{p}_b^2 - 1) = \frac{1}{2} (V_{11}^2 + V_{33}^2 - 1). \\ 
    \end{aligned}
\end{equation}

For $\xi=1.0$ and $\theta=\pi/4$, we have

\begin{equation}
\begin{aligned}
\mathrm{det}(D)_{ \{ 1, 3\}} &= \begin{vmatrix}
\hat{a}^\dagger\hat{a} & \hat{a}^\dagger\hat{b}^\dagger\\
\hat{a}\hat{b} & \hat{b}^\dagger \hat{b}
\end{vmatrix} \\
&= \begin{vmatrix}
\sinh^2(\xi)\cos^2(\theta)& \cosh(\xi)\sinh(\xi)\cos(\theta)\sin(\theta) \\
\cosh(\xi)\sinh(\xi)\cos(\theta)\sin(\theta) & \sinh^2(\xi)\sin^2(\theta) 
\end{vmatrix} \\
&= -0.3453
\end{aligned}
\end{equation}

\subsection*{Entangled cat states}
A single mode cat state can be described as a superposition of two coherent states, whose displacements are parametrized by their complex displacement $\alpha$, 

$$\ket{\psi_{cat_\pm}} = \frac{\ket{\alpha} \pm \ket{-\alpha}}{\sqrt{2\pm2e^{-2|\alpha|^2}}}.$$

To construct entangled cat states, two modes are required. On the first mode, we start off with a simple odd cat state $\ket{\psi_{cat_-}}$. The second mode is vacuum $\ket{\text{vac}}$, and the full input state is thus $\ket{\psi_{in}} = \ket{\psi_{cat_-}} \ \otimes \ket{\text{vac}}$. When running both modes through a beamsplitter, the effect on the creation operators is the following:

\begin{align}
\hat{a}^\dagger \rightarrow \frac{1}{\sqrt{2}}\hat{a}^\dagger + \frac{1}{\sqrt{2}}b^\dagger, \\
\hat{b}^\dagger \rightarrow \frac{1}{\sqrt{2}}\hat{a}^\dagger - \frac{1}{\sqrt{2}}\hat{b}^\dagger.
\end{align}

Each coherent state of the cat mode can be treated separately when mixing with the vacuum mode on the beamsplitter,
\begin{align}
\ket{\alpha}_a \otimes \ket{\text{vac}}_b &\rightarrow \left|\frac{\alpha}{\sqrt{2}}\right\rangle_a \otimes \left|\frac{\alpha}{\sqrt{2}}\right\rangle_b,  \\
\ket{-\alpha}_a \otimes \ket{\text{vac}}_b &\rightarrow \left|-\frac{\alpha}{\sqrt{2}}\right\rangle_a \otimes \left|-\frac{\alpha}{\sqrt{2}}\right\rangle_b.
\end{align}

Finally, we have our entangled cat state
$$\ket{\psi_{ent}} = \frac{1}{\sqrt{2-2e^{-2|\alpha|^2}}}\left( \left| \frac{\alpha}{\sqrt{2}}, \frac{\alpha}{\sqrt{2}}\right\rangle - \left| -\frac{\alpha}{\sqrt{2}}, -\frac{\alpha}{\sqrt{2}}\right\rangle\right).$$

Expectation values are then easy to derive. One important equation to keep in mind when calculating the overlap between two coherent states is that the coherent state basis is overcomplete due to $\braket{\beta| \alpha} = e^{-\frac{1}{2}|\alpha-\beta|^2 +\frac{1}{2}(\alpha \beta^* - \alpha^*\beta)}$. We can define the overlaps between opposite-sign coherent states as

\begin{equation}
\braket{\alpha | -\alpha} = \braket{-\alpha | \alpha} = e^{-2|\alpha|^2}
\end{equation}

Finally, if we redefine $\alpha/\sqrt{2}$ as $\alpha'$, we can derive the expectation values of relevant operators:
\begin{align}
\langle \hat{b}^\dagger \hat{b}\rangle = \langle \hat{a}^\dagger \hat{b}\rangle = \langle \hat{a} \hat{b}^\dagger \rangle &= \frac{|\alpha'|^2( 1 + e^{-|2\alpha'|^2})}{(1-e^{-|2\alpha'|^2})} \\ 
\langle \hat{a}^\dagger \hat{a}\hat{b}^\dagger \hat{b} \rangle &= |\alpha'|^4 \\
\langle \hat{b}^\dagger \rangle = \langle \hat{b} \rangle &= 0
\end{align}

\subsection*{Photon-subtracted state}
We reproduce almost verbatim Serafini's \cite{Serafini2021-pd} thorough description of the photon-subtracted state. If we assume a two-mode squeezed vacuum (TMSV) in modes $a$ and $b$, the state in the Fock basis can written as

\begin{equation}
\ket{\psi_{\mathrm{TMSV}}} = \text{cosh}(r)^{-1}\sum_{n=0}^\infty \text{tanh}(r)^n\ket{n, n}.
\end{equation}

Now, introducing ancilla modes $c$ and $d$ on which the photon detection will occur, we have the full state

\begin{equation}
\ket{\psi} = \ket{\psi_{\mathrm{TMSV}}} \otimes \ket{0, 0} = \sum_{n=0}^\infty \frac{\tanh(r)^n}{\text{cosh}(r)}\ket{n, n, 0, 0}.
\end{equation}

After mixing modes pairs ($a, c$) and ($b, d$) using beamsplitters of transmittivity $\theta$, we can transform the above state with the beamsplitter operator $\hat{R}_\theta^{(ac)\dagger}$, 

\begin{equation}
\label{eq:photon_subtracted_state_exact}
\begin{aligned}
    \ket{\psi_{\mathrm{out}}} &= \hat{R}_\theta^{(ac)\dagger}\hat{R}_\theta^{(bd)\dagger}(\ket{\psi_{\mathrm{TMSV}}} \otimes \ket{0, 0})\\
    &=\sum_{n=0}^\infty\sum_{j=0}^n\sum_{k=0}^n \frac{[\text{cos}^2\theta \,\text{tanh}(r)]^n}{\text{cosh}(r)}(-\text{tanh}|\theta|)^{j+k} \frac{\hat{a}^j \hat{b}^k}{\sqrt{j! k!}}\ket{n, n, j, k}.
\end{aligned}
\end{equation}

If we Taylor expand the above around $\theta$ using $\cos^2\theta = 1 - \theta^2 + \mathcal{O}(\theta^2)$ and $\tan\theta = \theta + \mathcal{O}(\theta^2)$, we get the following approximation for $\theta \ll 1$:

\begin{equation}
\ket{\psi_{\mathrm{out}}} = \left(\ket{\psi_{\mathrm{TMSV}}} - \theta^2\sum_{n=0}^\infty\frac{n\tanh(r)^n}{\cosh(r)}\ket{n, n}\right)\otimes\ket{0,0} - \theta\hat{a}\ket{\psi_{\mathrm{TMSV}}}\otimes\ket{1, 0} - \theta\hat{b}\ket{\psi_{\mathrm{TMSV}}}\otimes\ket{0, 1} + \theta^2\hat{a}\hat{b}\ket{\psi_\mathrm{TMSV}}\otimes\ket{1, 1} + \mathcal{O}(\theta^2).
\end{equation}

If we herald our experiment on the successful detection of a single photon at both photodetectors, the fourth term in the sum is the leading-order for the heralded output. The success probability is

\begin{equation}
\begin{aligned}
P_{\mathrm{success}} &= \theta^4\bra{\psi_{\mathrm{TMSV}}}\hat{a}^\dagger\hat{b}^\dagger\hat{a}\hat{b}\ket{\psi_{\mathrm{TMSV}}} + O(\theta^4)\\
&= \theta^4\cosh^{-2}(r) \sum_{n=0}^\infty \tanh^{2n}(r) n^2 + O(\theta^4) \\
&\overset{(\text{i})}{=}  \theta^4\cosh^{-2}(r) \frac{\tanh^2(r)(1+\tanh^2(r))}{(1 - \tanh^2(r))^3}  + O(\theta^4) \\
&\overset{(\text{ii})}{=} \theta^4 \tanh^2(r)(1+\tanh^2(r))\cosh^4(r) + O(\theta^4)\\
&\overset{(\text{iii})}{=} \theta^4 \sinh^2(r)(2\sinh^2(r) + 1) + O(\theta^4)
\end{aligned}
\end{equation}

\noindent where in $(i)$ we have used the relation $\sum_{n=1}^\infty n^2 x^{2n} = -\frac{x^2 (1 + x^2)}{(-1 + x^2)^3}$ and in $(ii)$, $(iii)$ we have used the hyperbolic relations $\cosh^2(r) - \sinh^2(r) = 1$ and $\tanh(r) = \sinh(r)/\cosh(r)$. The final photon-subtracted state is then

\begin{equation}
\ket{\psi_\mathrm{sub}} = \frac{\hat{a}\hat{b}\ket{\psi_{\mathrm{TMSV}}}}{\sinh(r)\sqrt{2\sinh(r)^2+1}} = \sum_{n=0}^\infty \frac{2\tanh(r)^{n+1}(n+1)\ket{n, n}}{\sinh(2r)\sqrt{2\sinh(r)^2+1}} \ \ \ .
\end{equation}

This concludes the description of the photon-subtracted as presented in \cite{Serafini2021-pd}. Next, we can derive the required expectation values for completing the SV submatrices using Eq. \eqref{eq:photon_subtracted_state_exact}. If we postselect the full state such that we have exactly one photon in the ancilla modes, we are left with a simple sum over all possible Fock states in the two main modes of the photon-subtracted state. We denote by $C(n, \theta, r) \equiv \frac{\left[\cos^2(\theta) \tanh(r)\right]^{n+1}}{\cosh(r)} \tanh(|\theta|)^2 (n+1)$ the coefficients for a Fock state $\ket{n, n}$ after postselection, as described in Eq. \eqref{eq:photon_subtracted_state_exact}. We therefore have as a state after postselection

\begin{equation}
\label{eq:postselected_state}
\begin{aligned}
\ket{\psi_\mathrm{sub}} = \frac{1}{\gamma} \sum^\infty_{n=0}C(n, \theta, r) \ket{n, n}
\end{aligned}
\end{equation}

\noindent where $\gamma = \sqrt{\sum^\infty_{n=0} C^2(n, \theta, r)}$ is the renormalization factor after postselection. Deriving the expectation values is as simple as plugging Eq. \eqref{eq:postselected_state} into the calculation of the expectation value

\begin{equation}
\label{eq:photon_subtracted_expectations}
\begin{aligned}
\langle \hat{a}^\dagger \hat{a}\rangle = \langle \hat{b}^\dagger \hat{b}\rangle&= \frac{1}{\gamma^2} \left( \sum^\infty_{n=0} C(n, \theta, r) \bra{n, n} \right)\left( \sum^\infty_{n=0} n \, C(n, \theta, r) \ket{n, n} \right)\\
&= \frac{1}{\gamma^2} \sum^\infty_{n=0} n \, C^2(n, \theta, r)\\
&= \frac{
2\cos^4(\theta)\tanh^2(r)\left(2+\cos^4(\theta)\tanh^2(r)\right)
}{
1-\cos^8(\theta)\tanh^4(r)}\\
\langle \hat{a} \hat{b} \rangle = \langle \hat{a}^\dagger \hat{b}^\dagger \rangle &= \frac{1}{\gamma^2} \left( \sum^\infty_{n=0} C(n, \theta, r) \bra{n, n} \right)\left( \sum^\infty_{n=0} (n+1) \, C(n+1, \theta, r) \ket{n, n} \right)\\
&= \frac{1}{\gamma^2}\sum^\infty_{n=0} (n+1) C(n, \theta, r) C(n+1, \theta, r)\\
&= \frac{
2\cos^2(\theta)\tanh(r)\left(1+2\cos^4(\theta)\tanh^2(r)\right)
}{
1-\cos^8(\theta)\tanh^4(r)
}
\end{aligned}
\end{equation}

We can now try to derive the SV determinant for different values of transmittivity $\theta$ and squeezing $r$.

\end{document}